# Data-driven Augmentation of a Turbulence Model in Three-dimensional Separated Flows

Chenyu Wu[†], Shaoguang Zhang[‡], Yufei Zhang[*, ††, §]

School of Aerospace Engineering, Tsinghua University, Beijing, China, 100084

**Abstract**

Classic turbulence models often struggle to accurately predict complex flows. Although data-driven techniques have addressed these shortcomings, most existing research has concentrated on two-dimensional (2D) cases. This study bridges this gap by enhancing a data-driven turbulence model, the SST-CND (shear stress transport-conditioned) model, which was originally trained on 2D separated flows, in 3D scenarios. An additional correction term, $\beta_{3D}$, is introduced to account for 3D effects. The distribution of this term is determined through a 3D field inversion process using high-fidelity data obtained from the flow around a cube. An algebraic expression for $\beta_{3D}$ is then derived through symbolic regression and formulated to degrade to zero in 2D cases. The performance of the resulting SST-CND3D model is evaluated across a range of flows. In 2D flows, the SST-CND3D model performs identically to its 2D-trained predecessor. However, the model exhibits superior performance in 3D flows, such as the flow around the complex JAXA standard model high-lift configuration. These findings indicate that a sequential approach, constructing a 3D correction term that vanishes in 2D on top of a 2D-trained model, constitutes a promising method for developing data-driven turbulence models that perform accurately in 3D while preserving their effectiveness in 2D.

# 1. Introduction

Turbulence occurs in numerous engineering applications and can significantly affect the performance of engineering designs, making the analysis of turbulence highly important. In computational fluid dynamics (CFD), researchers often employ turbulence simulation methods to analyze turbulent flows. Among these methods, the Reynolds-averaged Navier-Stokes (RANS) equations are widely used in real-world engineering applications due to their high efficiency and low computational cost. However, evidence has shown that traditional RANS turbulence models often fail in complex flow conditions, such as flows involving separation [1][2][3] and flows around high-lift devices [4][5]. Duraisamy et al. [6] predicted that the RANS method will continue to serve as the workhorse of CFD in the future due to the immature state of scale-resolving methods. Therefore, considering the broad applicability of the RANS method, it is essential to enhance the accuracy of RANS turbulence models in complex flows.

[†] Ph.D. student, School of Aerospace Engineering, Tsinghua University

[‡] Post-doc researcher, School of Aerospace Engineering, Tsinghua University

[*] Associate professor, School of Aerospace Engineering, Tsinghua University (Corresponding author: zhangyufei@tsinghua.edu.cn)

[††] Associate professor, State Key Laboratory of Advanced Space Propulsion, Tsinghua University, 100084, Beijing

[§] Associate professor, Caofeidian Laboratory, 063200, Tangshan, China

Recently, researchers have increasingly used data-driven techniques to improve the ability of RANS turbulence models to predict complex separated flows [7]. Several studies have focused on improving the expression of Reynolds stress. Ling et al. [8] proposed a tensor-basis neural network to predict the nonlinear Reynolds stress, enhancing the model's accuracy in separated flows. Yin et al. [9] developed a novel feature decomposition method to increase the smoothness of the Reynolds stress predicted by the neural network. Tang et al. [10] and Zhao et al. [11] applied symbolic learning methods to construct analytical expressions for nonlinear Reynolds stress based on high-fidelity data. Rincón et al. [12] used a progressive learning approach, which corrects the Reynolds stress anisotropy first and then considers the flow separation corrections. The method was proven to be effective. Another group of studies has focused on correcting the transport equations of turbulence models. Singh et al. [13] introduced the field inversion and machine learning (FIML) method to quantify functional errors in the transport equations caused by assumptions made by modelers using high-fidelity data. Holland et al. [14] developed a variant of the FIML method that trains the neural network directly using the gradient produced by field inversion, achieving physics-consistent training. Srivastava et al. [15] proposed a sequential learning approach to learn two correction terms ($\beta_1$ and $\beta_2$) of the transition model's $\gamma$ equation. The effects of them are isolated by a novel blending function. The result showed that the proposed sequential learning approach can effectively predict bypass transition with and without flow separation, showing decent generalizability. Wu et al. [16] employed a symbolic regression method to train an analytical expression for the correction term based on field inversion data, resulting in a more interpretable FIML model with strong generalizability. Zhang et al. [17] applied the data-driven SST-CND (shear-stress-transport-conditioned) model trained by Wu et al. [18] using FIML to three-dimensional complex high-lift configurations. The results demonstrated that the SST-CND model performed better than the SST model but still showed some deviation from experimental data. Zhang et al. attributed this discrepancy to the lack of three-dimensional (3D) data in the training set of the SST-CND model, which was trained only on two-dimensional (2D) cases.

Most studies on data-driven turbulence models train models using 2D high-fidelity data, with only a few utilizing 3D data. Ho et al. [19] performed FIML on a mixed dataset that included field inversion data from a 3D hill. The machine learning model developed by Ho et al. was tested on a 3D non-axisymmetric bump, yielding favorable results. Yan et al. [20] conducted FIML on the axisymmetric 3D FAITH hill [21], analyzing the effect of the number of samples used in machine learning on the accuracy of the trained neural network. Sun et al. [22] trained a data-driven turbulence model based on neural networks using 3D wing data at high Reynolds numbers. The model was deployed on a heterogeneous supercomputing environment, demonstrating both high accuracy and efficiency. Steiner et al. [23] focused on 3D wind farms and derived analytical correction terms for the RANS equations using large eddy simulation data through symbolic regression. In all the studies mentioned above, 3D data were employed to develop data-driven turbulence models. However, all correction terms were derived from scratch without considering their impact on 2D flows, which can compromise the baseline model's accuracy in 2D flows and limit the model's generalizability.

This study uses a sequential correction approach to enhance the performance of the SST-CND model in three-dimensional (3D) complex flows using 3D data. The method involves constructing an additional correction term, $\beta_{3D}$, directly on top of the SST-CND model's original correction term ($\beta_{CND}$). The distribution of $\beta_{3D}$ is first obtained through field inversion of the flow around a cube. Symbolic regression is then employed to construct an explicit expression passing through the origin for $\beta_{3D}$ as a function of local flow features that are non-zero only in 3D flows. This deliberate design ensures that the $\beta_{3D}$ correction is inactive in two-dimensional (2D) cases, which is crucial because the original SST-CND model already performs well in these scenarios, and maintaining that performance is essential. The resulting SST-CND3D model exhibits identical performance to its predecessor on various 2D test cases while demonstrating superior accuracy in 3D complex high-lift configurations, such as the JAXA standard model. This study indicates that constructing a 3D correction term on top of a model trained exclusively on 2D data is a logical and effective approach to enhancing a model's 3D performance without compromising its baseline accuracy in 2D flows.

# 2. Methods

In this section, the model on which the 3D correction term is developed, namely the SST-CND model, is first introduced. Then, the field inversion method and the symbolic regression method utilized to construct the $\beta_{3D}$ correction term are briefly discussed.

## 2.1. Formulation of the SST-CND model and the $\beta_{3D}$ correction term

This study aims to enhance the performance of the SST-CND [18] model in 3D complex flows. The SST-CND model is a data-driven turbulence model trained using conditioned field inversion and symbolic regression. It performs effectively in predicting a series of 2D complex separated flows, including the NASA hump and NLR-7301 multi-element airfoil. It also maintains the original SST (shear-stress-transport) model's accuracy in simple wall-attached flows, such as the zero-pressure-gradient turbulent flat plate. The formulation of the SST-CND model is presented below (for incompressible flow with constant density).

$$
\begin{aligned}
&\frac{Dk}{Dt} = P_k - \beta^* k\omega + \nabla \cdot [(\nu + \sigma_k \nu_T)\nabla k] \\
&\frac{D\omega}{Dt} = \frac{\gamma}{\nu_T} P_k - \beta\theta\omega^2 + \nabla \cdot [(\nu + \sigma_\omega \nu_T)\nabla \omega] + \frac{2(1-F_1)\sigma_{\omega 2}}{\omega}\nabla k \cdot \nabla \omega
\end{aligned}
\tag{1}
$$

$k$ is the turbulence kinetic energy; $\omega$ is the specific dissipation rate; $\beta$ in Eq. (1) (appearing as a multiplier in front of the destruction term of the $\omega$ equation) is the correction term derived through conditioned field inversion and symbolic regression. The training data are entirely 2D, including the NASA hump case [24] and the curved backward-facing step case [25]. For details of the training process, please refer to Ref [18]. The formulation of $\beta$ is provided as follows:

$$
\begin{aligned}
&\beta = \beta_{CND} f_d + 1 \\
&\beta_{CND} = \min(0.00435\lambda_2^2, 3.80686) \\
&\lambda_2 = \mathrm{tr}\left[\left(\frac{\mathbf{\Omega}}{\beta^*\omega}\right)^2\right], \Omega_{ij} = \frac{1}{2}\left(\frac{\partial u_i}{\partial x_j} - \frac{\partial u_j}{\partial x_i}\right) \\
&f_d = 1 - \tanh[(8r_d)^3], r_d = \frac{\nu + \nu_T}{\kappa^2 d^2 \sqrt{\frac{\partial u_i}{\partial x_j}\frac{\partial u_i}{\partial x_j}}}
\end{aligned}
\tag{2}
$$

Note that if $\beta = 1$, the model degrades to the baseline SST model. The notations used in Eq. (3) are slightly different from the ones used in Ref [18], but the models presented in both papers are identical. The definitions of other terms, such as $P_k$ and $\theta$, and the value of the model constants, can be found in Ref [26].

The SST-CND model was successfully applied to 3D complex high-lift configurations [17] such as the JAXA Standard Model (JSM). The SST-CND model demonstrates better accuracy compared to the baseline SST model. However, the maximum lift coefficient ($C_L$) and the stall angle of attack ($AOA$) are still underpredicted by the SST-CND model. Zhang et al. [17] attributed this inaccuracy to the SST-CND model's lack of 3D training data and its inability to capture 3D effects. Therefore, adding 3D data to the training set can potentially improve the model's performance in complex 3D flows such as the JSM. However, retraining the model from scratch using a mixed dataset of 2D and 3D data is undesirable. The SST-CND model already performs well in a series of 2D separated flows, and retraining can compromise this established performance. Hence, a method is required to incorporate 3D training data non-intrusively, enhancing 3D performance while preserving the model's existing 2D accuracy.

This study addresses the aforementioned issue by introducing an additional correction term $\beta_{3D}$, into the SST-CND model and formulating an expression for $\beta_{3D}$ through a sequential learning approach:

$$\beta = (\beta_{CND} + \beta_{3D}) f_d + 1 \tag{3}$$

The spatial distribution of $\beta_{3D}$ is determined through field inversion of the flow around a cube (3D). Then, symbolic regression is applied to construct an expression for $\beta_{3D}$ that vanishes in 2D flows. In this manner, the 3D training data are injected exclusively into the $\beta_{3D}$ term, avoiding retraining of the $\beta_{CND}$ term. A $\beta_{3D}$ term that reduces to zero in 2D scenarios also ensures that the model's performance in 2D flows remains unaffected.

## 2.2. Field inversion

In this section, the field inversion method utilized to obtain the spatial distribution of $\beta_{3D}$ is introduced. During the field inversion process, the distribution of $\beta_{3D}$ is adjusted by an optimization algorithm to match the RANS-predicted velocity with the high-fidelity data (i.e., experimental data). The $\beta_{3D}$ distribution that minimizes the discrepancy between the RANS-predicted velocity and the high-fidelity data is defined as the optimal $\beta_{3D}$ distribution. We prefer the velocity probe data because we believe that it can better represent the spatial structure of the flow field, which is important for separated flow. In this study, it is obtained by solving the following specific optimization problem.

$$\min_{\boldsymbol{\beta}_{3D}} J = \lambda_{obs} \sum_i \left[u_i^{exp} - u_i(\boldsymbol{\beta}_{3D})\right]^2 + \lambda_{prior} \sum_j \beta_{3D,j}^2 \tag{4}$$

$\beta_{3D,j}$ is the value of $\beta_{3D}$ in the $j^{th}$ cell of the computational mesh; $\boldsymbol{\beta}_{3D}$ is the vector whose $j^{th}$ element is $\beta_{3D,j}$; $u_i^{exp}$ is the $i^{th}$ high-fidelity data point of streamwise velocity provided by the experiment; $u_i(\boldsymbol{\beta}_{3D})$ is the RANS-predicted streamwise velocity corresponding to the $i^{th}$ high-fidelity data point given the vector $\boldsymbol{\beta}_{3D}$. The optimal distribution of $\beta_{3D}$ corresponds to the optimal $\boldsymbol{\beta}_{3D}$ vector that can minimize $J$ in Eq. (4). $\lambda_{obs}$ and $\lambda_{prior}$ are two constants utilized to scale the two terms in Eq. (4). The first term in Eq. (4) measures the discrepancy between the RANS-predicted streamwise velocity and the high-fidelity data provided by the experiment. The second term in Eq. (4) measures the deviation of $\beta_{3D}$ from its default value, namely, zero.

A gradient-based optimization program, SNOPT [27], is employed to solve the optimization problem in Eq. (4). The discrete adjoint method [28] is utilized to compute the gradient of the objective function in Eq. (4) with respect to $\boldsymbol{\beta}_{3D}$. An introduction of the discrete adjoint method can be found in Ref. [29]. In this study, the open-source package DAFoam [30][31] is utilized to solve the discrete adjoint problem. The automatic differentiation in DAFoam is supported by the open-source package CODIPACK [32].

## 2.3. Symbolic regression

In this part, the symbolic regression (SR) method utilized to construct the analytical expression of $\beta_{3D}$ is briefly introduced. We selected Symbolic Regression (SR) as our learning algorithm primarily for two reasons. First, SR produces models in a closed-form analytical expression, which significantly streamlines both implementation and distribution. Secondly, the resulting expression exhibits inherently greater stability and robustness when extrapolating to new cases, as the functional relationship between the inputs and output is far simpler and more parsimonious than that of black-box alternatives like neural networks. The learning target of SR is the optimal distribution of $\beta_{3D}$ obtained through the field inversion process. As mentioned earlier, the expression of $\beta_{3D}$ should vanish in 2D flows. This requirement necessitates a careful selection of the input features used in symbolic regression. For 2D incompressible flows, the third and fourth invariants ($\lambda_3$ and $\lambda_4$) of the Reynolds stress are identically zero [33]. This can be proved by using the traceless property of $\hat{\boldsymbol{S}}$ and $\hat{\boldsymbol{\Omega}}$ in incompressible flows, and noticing that $\hat{\boldsymbol{S}}^2$ and $\hat{\boldsymbol{\Omega}}^2$ are proportional to the 2D identity tensor in 2D scenarios. This characteristic can be exploited to construct a series of features that are zero in 2D flows. The features constructed through the transformation of $\lambda_3, \lambda_4$, as well as those obtained by multiplying $\lambda_3, \lambda_4$ with other features, are listed in Table 1. All of these features are zero in 2D incompressible flows, enabling them to distinguish between 2D and 3D flows, allowing $\beta_{3D}$ to vanish in 2D cases. Some features were used in classical turbulence modeling, such as $|\lambda_4|$, which appears in the Wilcox $k-\omega$ 2006 model [34]. In contrast, other features are constructed artificially.

Table 1. Input features of symbolic regression.

| Name | Definition | Physical Meaning |
|---|---|---|
| $\lambda_3$ | $\mathrm{tr}(\hat{\boldsymbol{S}}^3)$ | The third and fourth independent invariants of tensors $\hat{\boldsymbol{S}}$ and $\hat{\boldsymbol{\Omega}}$, where $\hat{\boldsymbol{S}}=\frac{k}{\varepsilon}\boldsymbol{S}, \hat{\boldsymbol{\Omega}}=\frac{k}{\varepsilon}\boldsymbol{\Omega}, \varepsilon=\beta^{*}k\omega$. $\boldsymbol{S}$ and $\boldsymbol{\Omega}$ are the strain rate tensor and the rotation rate tensor. The invariants $\lambda_3$ and $\lambda_4$ are related to the strength and the structure of the vorticity in the flow field. They are theoretically zero in 2D incompressible flows. |
| $\lambda_4$ | $\mathrm{tr}(\hat{\boldsymbol{\Omega}}^2\hat{\boldsymbol{S}})$ | |
| $l_{14}$ | $\lambda_1\lambda_4, \lambda_1=\mathrm{tr}(\hat{\boldsymbol{S}}^2)$ | $\lambda_1$ is utilized to measure the strength of the strain rate in the flow. Multiplying $\lambda_1$ with $\lambda_4$ or $\lambda_3$ makes the feature zero in 2D flows. |
| $l_{13}$ | $\lambda_1\lambda_3, \lambda_1=\mathrm{tr}(\hat{\boldsymbol{S}}^2)$ | |
| $\lvert\lambda_3\rvert$ | $\lvert\mathrm{tr}(\hat{\boldsymbol{S}}^3)\rvert$ | The magnitude of $\lambda_3$ and $\lambda_4$. It is a simple transformation of the features. Note that $\lvert\lambda_4\rvert$ was used in the Wilcox $k-\omega$ turbulence model to distinguish the round jet [34]. |
| $\lvert\lambda_4\rvert$ | $\lvert\mathrm{tr}(\hat{\boldsymbol{\Omega}}^2\hat{\boldsymbol{S}})\rvert$ | |
| $\lvert\lambda_3\rvert\mathrm{Re}_{\Omega}$ | $\lvert\mathrm{tr}(\hat{\boldsymbol{S}}^3)\rvert\frac{\lvert\boldsymbol{\Omega}\rvert d^2}{\nu}$ | $\mathrm{Re}_{\Omega}$ is originally utilized to distinguish the separated shear layer from the wall. These two features are constructed artificially to be related to separated shear layers and to be 0 in 2D incompressible flow at the same time. |
| $\lvert\lambda_4\rvert\mathrm{Re}_{\Omega}$ | $\lvert\mathrm{tr}(\hat{\boldsymbol{\Omega}}^2\hat{\boldsymbol{S}})\rvert\frac{\lvert\boldsymbol{\Omega}\rvert d^2}{\nu}$ | |

In this study, the open-source SR software PySR [35] is employed to derive an analytical expression for $\beta_{3D}$. The element functions employed in the SR process are listed in Table 2. PySR selects the features in Table 1 and the functions in Table 2 to assemble the expression for $\beta_{3D}$. It utilizes an evolutionary algorithm with annealing to fit the expressions to the data. The L2 distance loss function (the default in PySR) is adopted to evaluate the fitting quality.

$$loss(E) = \sum_{j} \left[\beta_{3D,j} - y_j\right]^2 \tag{5}$$

$E$ Stands for an expression; $y_j$ is the expression's prediction of the $j^{th}$ data point; PySR also defines a metric $C(E)$ to measure the complexity of the expression $E$. This study determines $C(E)$ by:

$$C(E) = N_{opr.} + 4N_{const.} + 2N_{var.} \tag{6}$$

$N_{opr.}$ is the number of operators in $E$; $N_{const.}$ is the number of constants; $N_{var.}$ is the number of variables. The maximum allowed $C(E)$ is set to 16. A higher weight is assigned to the variables to encourage PySR to perform feature selection, while an even higher weight is set for the constants to prevent PySR from overfitting by using too many constants. During the SR process, PySR maintains the best expressions across various complexity levels, allowing the user to select the most suitable expression by balancing complexity and loss.

Table 2. List of element functions used in symbolic regression

| **Operator type** | **Operators** |
|---|---|
| Binary operators | $(\cdot) + (\cdot), (\cdot) - (\cdot), (\cdot) \times (\cdot), (\cdot) \div (\cdot), \min(\cdot,\cdot), \max(\cdot,\cdot)$ |
| Unary operators | $\exp(\cdot), \tanh(\cdot), \frac{1}{1+(\cdot)}, \frac{1}{(\cdot)}$ |

# 3. Model training

This section discusses the model training process, specifically the construction of the $\beta_{3D}$ term. First, field inversion is conducted on the flow around a cube to determine the optimal $\beta_{3D}$ distribution. Then, SR is applied to formulate the analytical expression of $\beta_{3D}$. The final model, which integrates the constructed $\beta_{3D}$ expression, is referred to as the SST-CND3D model.

## 3.1. The field inversion of the flow around a cube

The flow around a cube is analyzed in this part. This case was investigated experimentally by Martinuzzi et al. [36]. It should be noted that Ref [37] showed that the unsteady RANS computation demonstrates better accuracy compared to the steady RANS computation. This is because of the inherent unsteady nature of the flow around the surface-mounted cube, which includes vortex shedding [38]. However, our goal here is to build a model that can potentially take the 'unresolved unsteadiness' in steady RANS computations into account. So we have chosen this inherently unsteady flow as the training set. From the engineering point of view, the flow around a surface-mounted cube also resembles the flow around the slat brackets in the high-lift device. This is also a reason for the choice of this training case. The flow field exhibits symmetry; therefore, only half of the computational domain is considered. The computational mesh is shown in Figure 1(a). consisting of approximately half a million cells. The side view of the computational domain is presented in Figure 1(b). The height of the cube is $H$, and the channel height is $2H$, which aligns with the experimental setup [36]. $H$ is set to 1 m in this study. The computational domain is $4.5H$ wide and $14.5H$ long, with the cube positioned $3.5H$ downstream of the inlet. The origin of the coordinate system is located at the intersection of the frontal lower edge of the cube and the symmetry plane. The inlet velocity follows a well-developed channel flow profile, with a bulk velocity of $0.6\ m/s$. The Reynolds number, based on the bulk velocity and cube height, is set to $5 \times 10^4$, consistent with the dataset provided in Ref. [36]. Figure 2 shows the high-fidelity data points (dark green dots) on the symmetry plane used in the field inversion process. These points correspond to the experimental velocity profiles at three positions ($x = 0.5\ m, x = 2.5\ m, x = 4.0\ m$) on the symmetry plane, as obtained by Martinuzzi et al. [36].

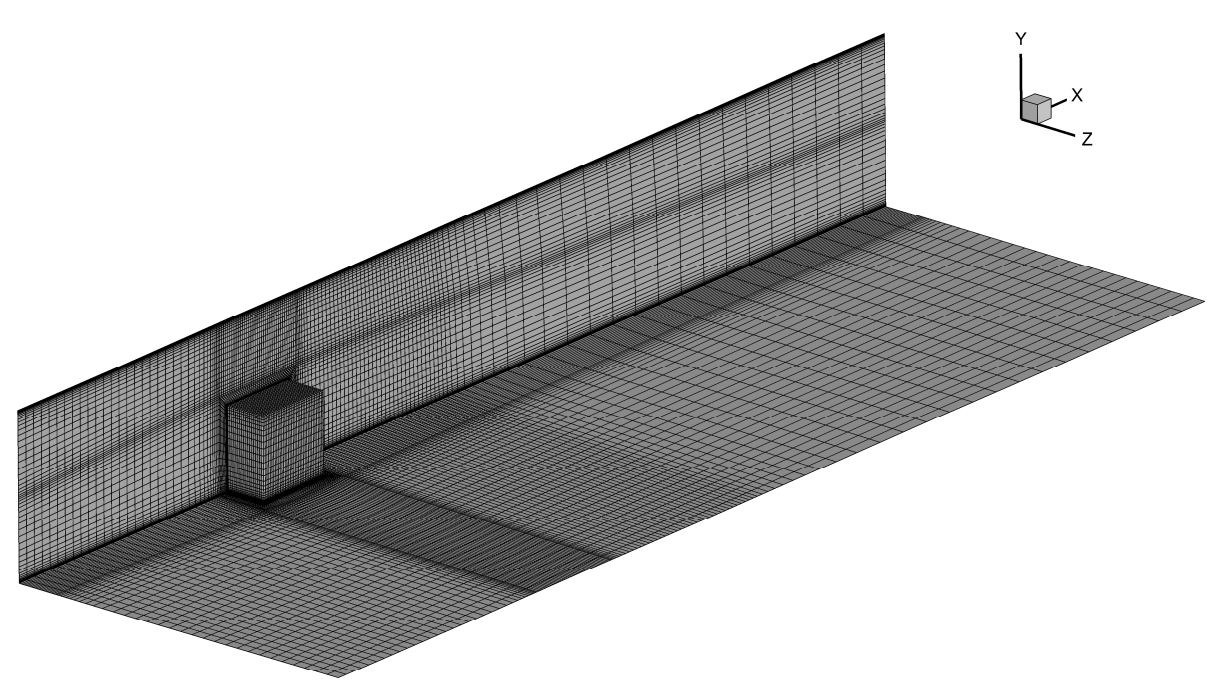

(a)

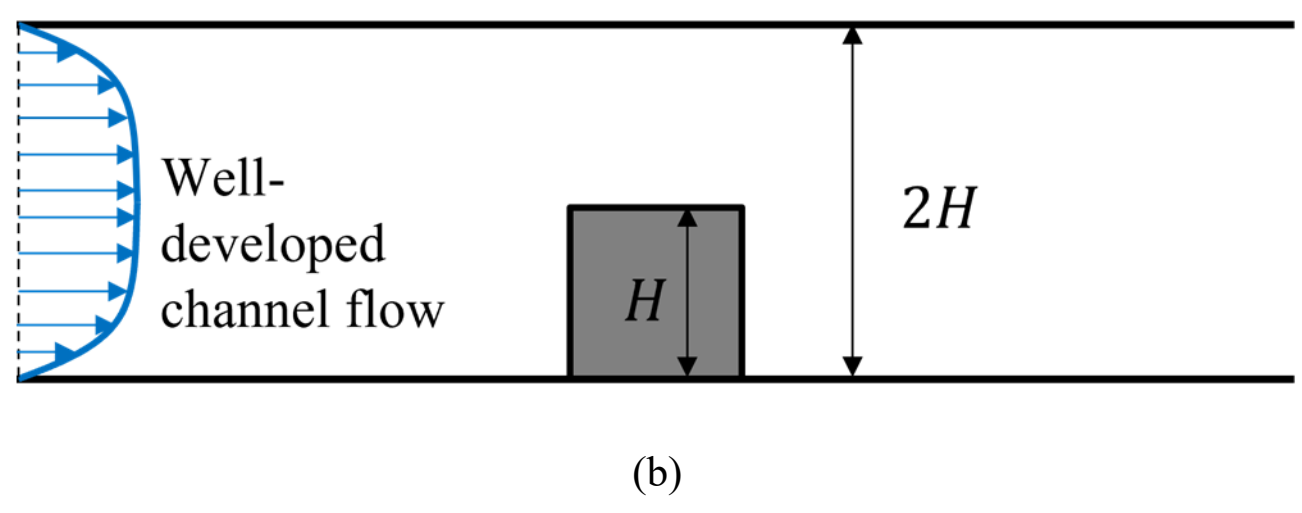

(b)

Figure 1. Computational domain of the cube case: (a) Computational mesh, (b) Side view of the domain

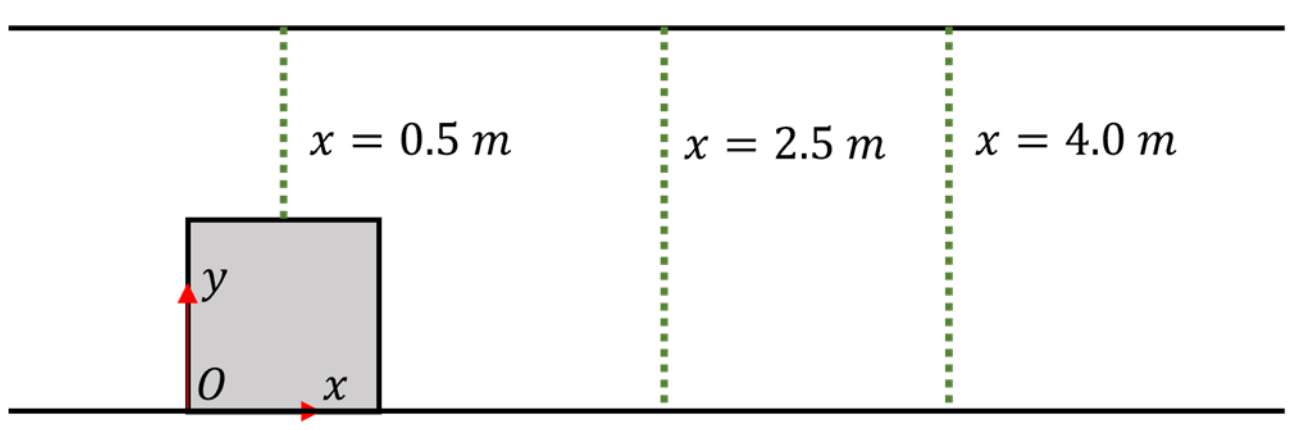

Figure 2. Location of the high-fidelity datapoints

The objective function used in the field inversion process is:

$$\min_{\boldsymbol{\beta}_{3D}} J = 1.0 \sum_i \left[u_i^{exp} - u_i(\boldsymbol{\beta}_{3D})\right]^2 + 1 \times 10^{-7} \sum_j \beta_{3D,j}^2 \tag{7}$$

The number of high-fidelity data points is approximately 100. The value of the regularization factor is set to $1 \times 10^{-7}$ and the upper bound of $\beta_{3D}$ is set to 8. The specific values are chosen by trial-and-error to reduce the error between the high-fidelity data and the prediction of RANS sufficiently. The optimization of the field inversion process runs for 225 steps (225 gradient evaluations), and the history is shown in Figure 3. The optimization approximately achieves convergence. The field inversion was run using 12 processors (Intel Xeon 4210R). It took around 500 seconds for the primal solution to meet the convergence criteria (residual $< 10^{-6}$, the number of iterations is around 500). A single adjoint run took around 2800 seconds to converge (residual $< 10^{-7}$). As a reference, the 2D turbulent flat plate boundary layer case in section 4.1 ran for around 2 minutes to iterate 8000 steps. Figure 4(a) indicates that the velocity profiles after the field inversion process match the high-fidelity data much better compared to the baseline SST-CND model (the model without $\beta_{3D}$ augmentation but with the $\beta_{CND}$ term). Figure 4(b) shows that the pressure distribution after the field inversion agrees better with the experiment [36]. The reference velocity for $C_p$ is taken as the bulk velocity. The experimental data are adjusted to ensure $C_p = 0$ at the outlet, matching the CFD settings. The results of the baseline SST model (without any correction term) in Figure 4 are also plotted for reference. Figure 5(a) shows the contour of $\beta_{3D}$ on the symmetry plane. $\beta_{3D}$ is mainly activated in the separated shear layer starting from the frontal edge of the cube and the region immediately in front of the cube. Figure 5(b) and Figure 5(c) indicate that the $\beta_{3D}$ correction term is also activated in the separated shear layer originating from the side edge of the cube. Figure 6 shows the contour of $\beta_{3D} = 7.99$, which can be viewed as the boundary of the region where $\beta_{3D}$ hits the upper limit 8. It can be seen that the contour only encloses a 'thin' region. Quantitative analysis further shows that the cells with $\beta_{3D} > 7.99$ only takes up 6% of the cells in the bounding box enclosing the whole separated flow region: $(x \times y \times z)$:$[-1.5\ m,\ 5.5\ m] \times [0.2\ m, 1.8\ m] \times [0.0\ m, 1.5\ m]$. Consequently, the choice of the upper bound of $\beta_{3D}$ can be viewed as reasonable.

The effect and the distribution of the $\beta_{3D}$ are physically interpretable. First, it can be found that an increased $\beta_{3D}$ means more destruction of $\omega$, which will decrease the value of $\omega$, thus suppressing the dissipation of $k$, which will finally result in an increased $\nu_T$. The increased $\nu_T$ brings a stronger momentum diffusion from the main stream to the separated region, which makes the size of the separation zone shrink. The increased $\nu_T$ given by the increased $\beta_{3D}$ in field inversion is clearly demonstrated in Figure 7. As for the distribution of $\beta_{3D}$, it coincides with the separated shear layer. This matches the observation of Rumsey[1] that the RANS turbulence models tend to underestimate the shear ($\nu_T$, in the framework of linear eddy viscosity models) in the separated shear layer. The possible 3D interference between separated shear layers, the spanwise flow, and the flapping of the separated shear layer further necessitates the increase in the $\beta_{3D}$ term to depict the strong momentum diffusion caused by these phenomena.

It should be noted that although only the symmetry plane data is used for field inversion, the correction field is enough to push the off-plane velocity profiles closer to the experiment. Figure 8 shows the off-plane velocity profiles, which indicate that the flow field after field inversion still matches the experimental data to a reasonable degree. However, in the present work, using only the symmetry plane data should be viewed as a simplification. This is because 3-D structures that exist in off-plane regions still exist and cannot be fully covered by the symmetry plane data. Though the current symmetry plane data seems to be enough to drive the off-plane velocity distribution to a reasonable degree, incorporating off-plane data in the field inversion process can potentially improve the inferred correction field.

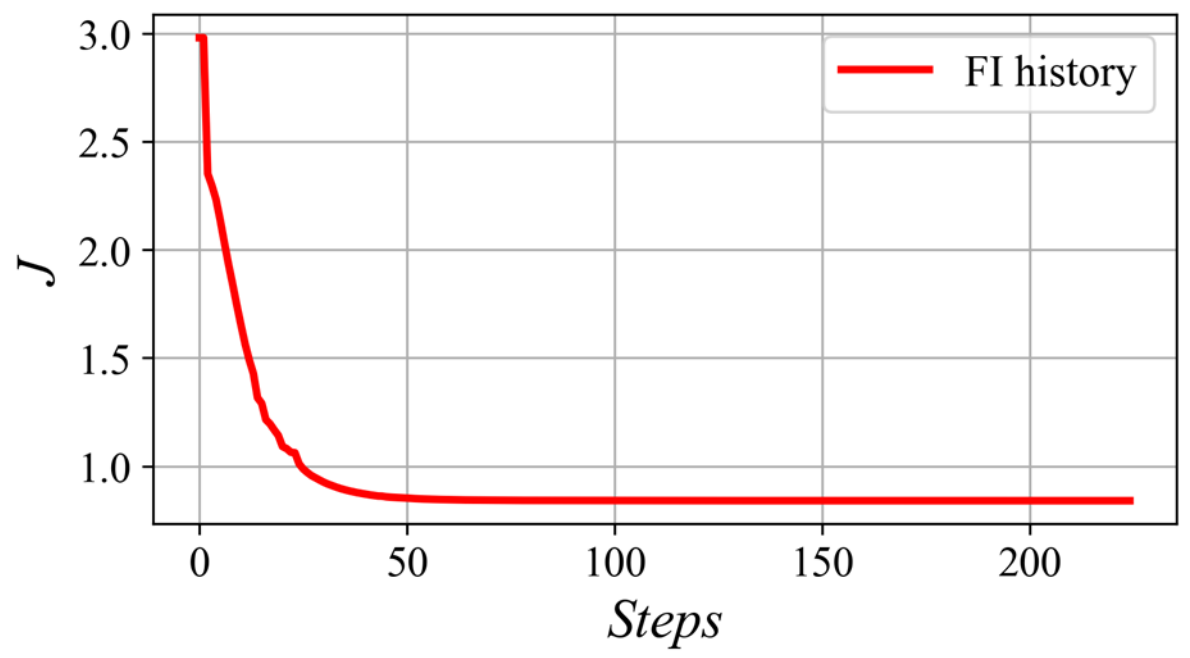

Figure 3. Convergence history of the objective function $J$

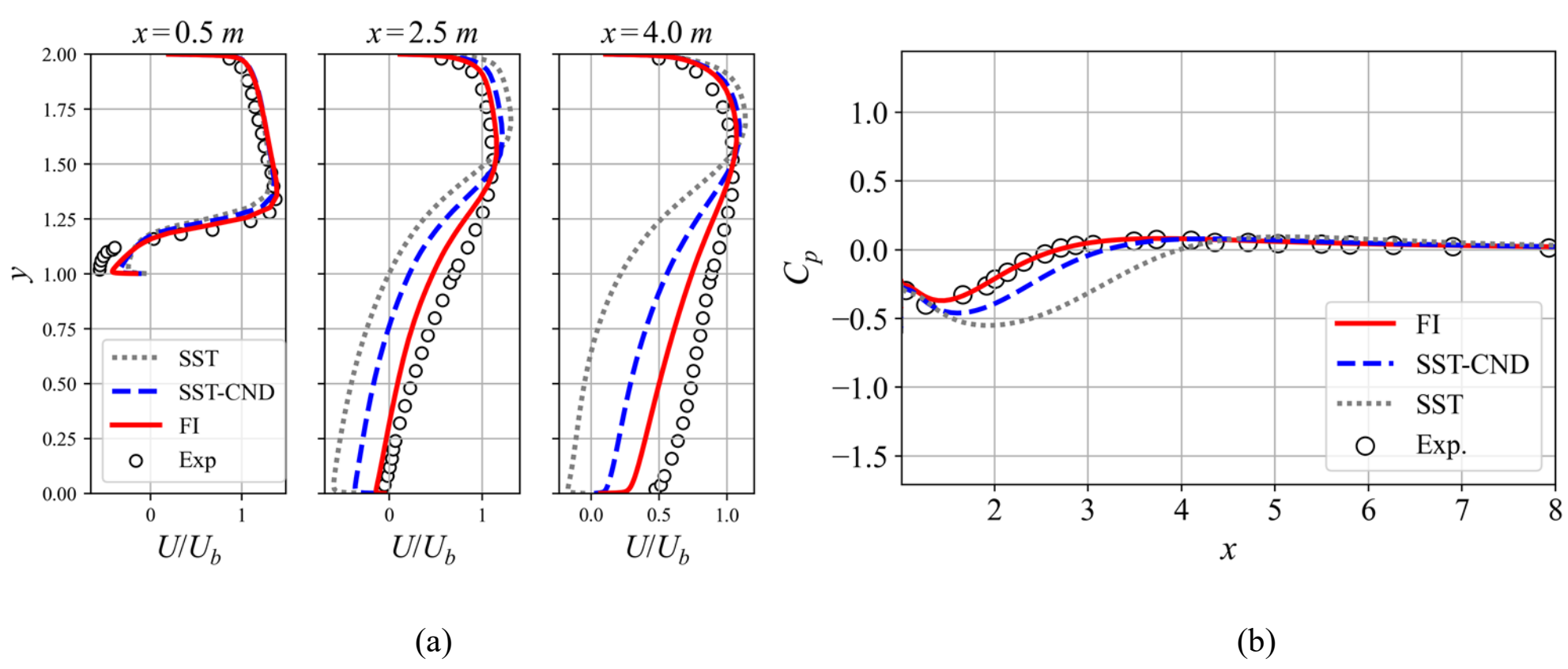

Figure 4. Flowfield after the field inversion, (a) Velocity profiles, (b) $C_p$ distribution

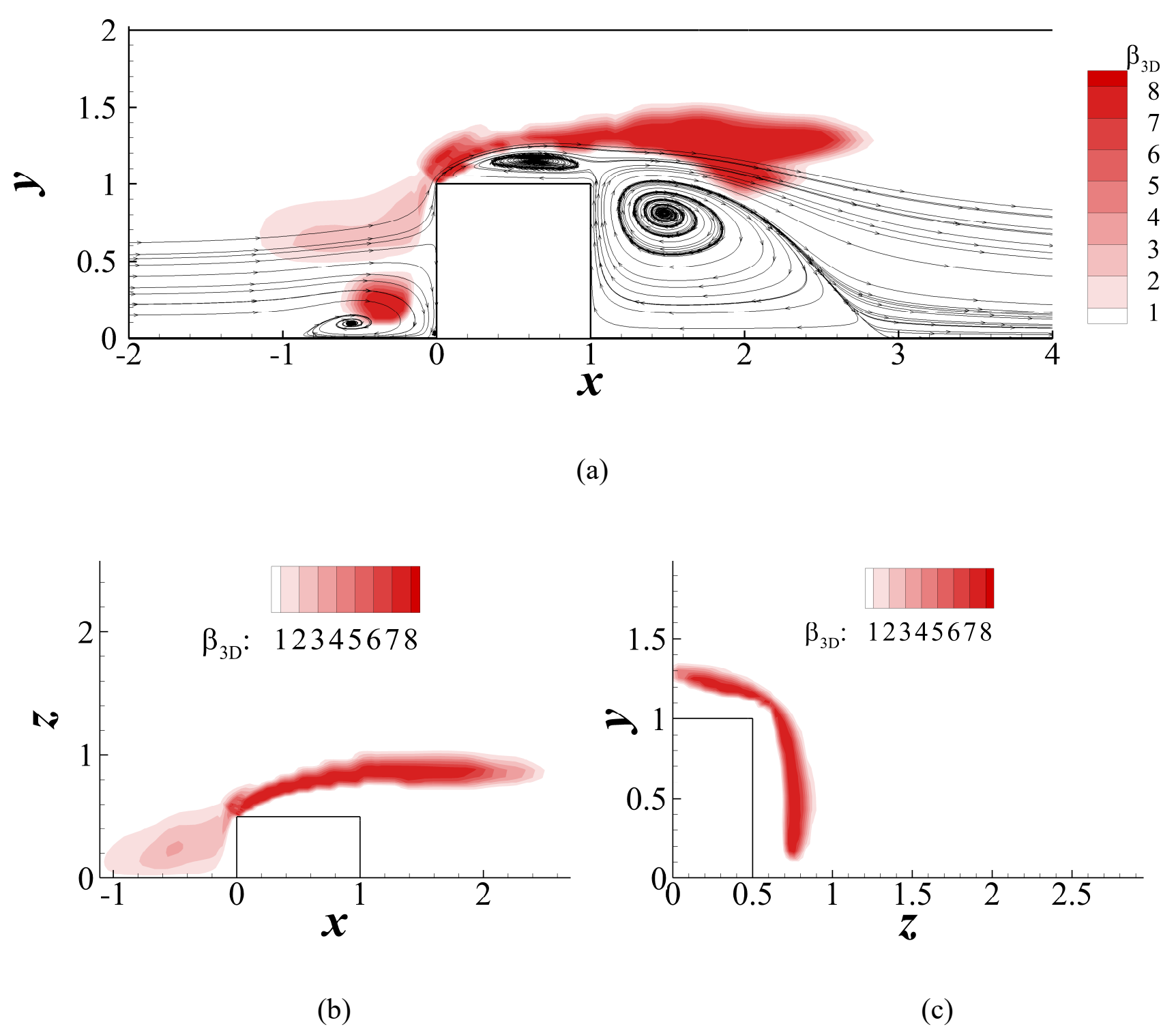

Figure 5. Contour of $\beta_{3D}$ after field inversion, (a) on the symmetry plane ($z = 0$), (b) on the $y = 0.5\,m$ plane, (c) on the $x = 0.5\,m$ plane

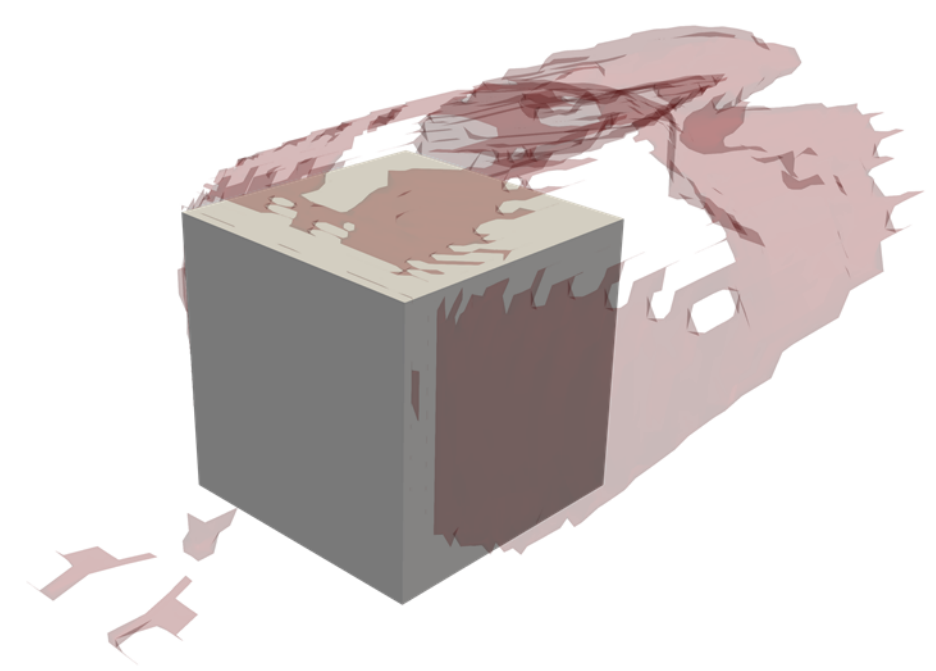

Figure 6. The 3D contour of $\beta_{3D} = 7{,}99$

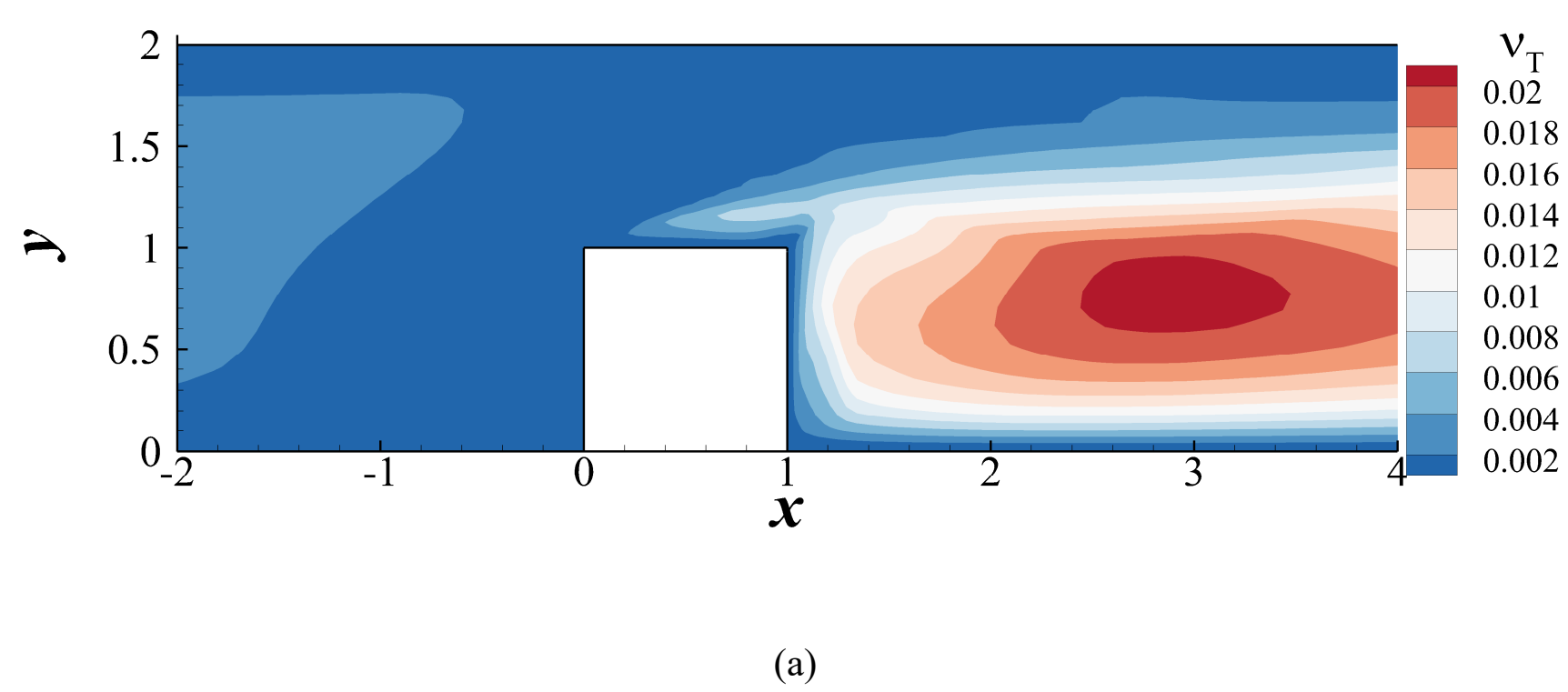

(a)

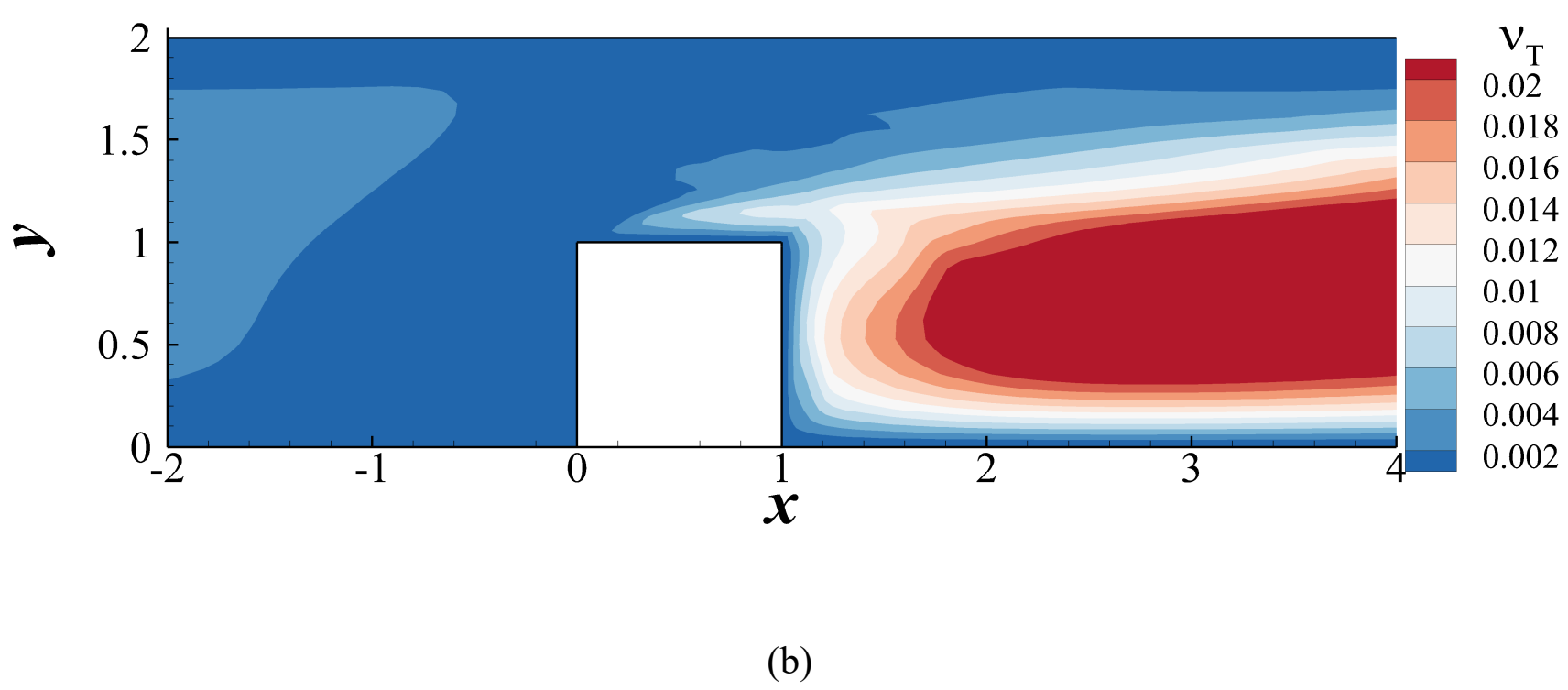

(b)

Figure 7. The $\nu_T$ contour at the symmetry plane given by (a) the SST-CND model, (b) the field inversion

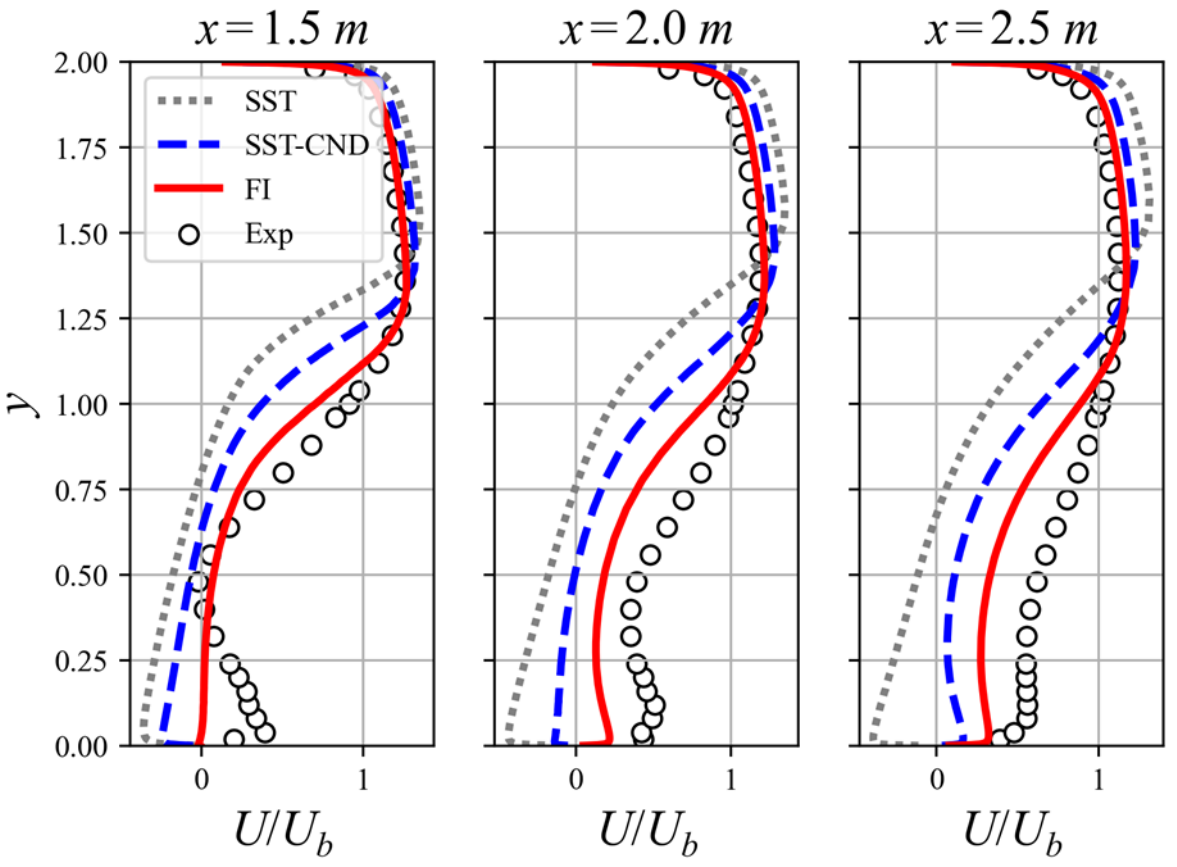

Figure 8. The velocity profiles on the $z = 0.5\ m$ plane

## 3.2. Constructing the expression of the $\boldsymbol{\beta_{3D}}$ term

The expression of the $\beta_{3D}$ term is constructed using symbolic regression based on the field inversion data. For the construction of the training dataset, all the cells in the bounding box ($x \times y \times z$): $[-1.5\ m,\ 5.5\ m] \times [0.2\ m, 1.8\ m] \times [0.0\ m, 1.5\ m]$ are first extracted. The total number of cells in this bounding box is approximately $1 \times 10^5$, which is too large for symbolic regression. Therefore, random down-sampling is performed among these cells to select only $1 \times 10^4$ of them as the training dataset. Using the features listed in Table 1 and the element functions in Table 2, PySR obtained a series of expressions with varying complexities. The final expressions are shown in Table 3. The loss of the expressions against the complexity is also illustrated in Figure 9. The loss decreases only slightly when the complexity exceeds 12, while extra features and complex functional structures are introduced. It is also required that the expression yields zero output in 2D flows (when the input features are zero) to switch off the augmentation in 2D scenarios. Based on the complexity-loss relationship and this requirement, the expression with a complexity of 12 is selected as the expression for $\beta_{3D}$:

$$\beta_{3D} = \min(0.38399717|\lambda_4|, 7.041212) \tag{8}$$

This produces $\beta_{3D} = 0$ in 2D flows since $|\lambda_4| = 0$ in such cases. The expressions with complexity of 15 and 16 are also tested in the training case (the cube case), and we find that they add little value compared to the expression with complexity of 12. So the final choice of the expression (complexity of 12) is reasonable. Eq. (8) is then integrated into the incompressible solver SimpleFoam of OpenFOAM [39]. It is also implemented in the compressible solver CFL3D [40]. In CFL3D, when computing $\lambda_4$, the correction proposed in Ref. [34] is utilized:

$$\lambda_4 = \mathrm{tr}\left(\widehat{\boldsymbol{\Omega}}^2\left(\widehat{\boldsymbol{S}} - q\boldsymbol{I}\right)\right), q = 0.5\mathrm{tr}(\widehat{\boldsymbol{S}}) \tag{9}$$

$\boldsymbol{I}$ is the identity tensor. This correction ensures that $\lambda_4$ is also identically 0 in 2D compressible flows since $\widehat{\boldsymbol{S}} - q\boldsymbol{I}$ is traceless for any 2D scenarios. Note that this correction has no effect for incompressible flows because $q = 0$ in such flows. The model integrated with the $\beta_{3D}$ term is named the SST-CND3D model.

Table 3. The final expressions obtained by PySR

| Complexity | Loss | Expression |
|---|---|---|
| 2 | 35.63232 | $\vert\lambda_4\vert$ |
| 3 | 4.2659435 | $\tanh(\vert\lambda_4\vert)$ |
| 5 | 4.1934953 | $\exp(\frac{1}{1+\frac{1}{\vert\lambda_4\vert}})$ |
| 7 | 2.741825 | $0.22195499\vert\lambda_4\vert$ |
| 8 | 2.7418249 | $\tanh(0.22568963)\ \vert\lambda_4\vert$ |
| 9 | 2.4547632 | $\frac{1}{1+\frac{1}{\vert\lambda_4\vert}-0.8714446}$ |
| **12** | **2.0753813** | $\mathbf{\min(0.38399717\vert\lambda_4\vert, 7.041212)}$ |
| 15 | 2.0627658 | $\min(0.39465824\min(\vert\lambda_4\vert, \vert\lambda_3\vert), 7.0523863)$ |
| 16 | 1.9694929 | $\min(\frac{1}{1+\exp\left(\frac{1}{1+\vert\lambda_3\vert+\lambda_3}\right)}\vert\lambda_4\vert, 6.8211412)$ |

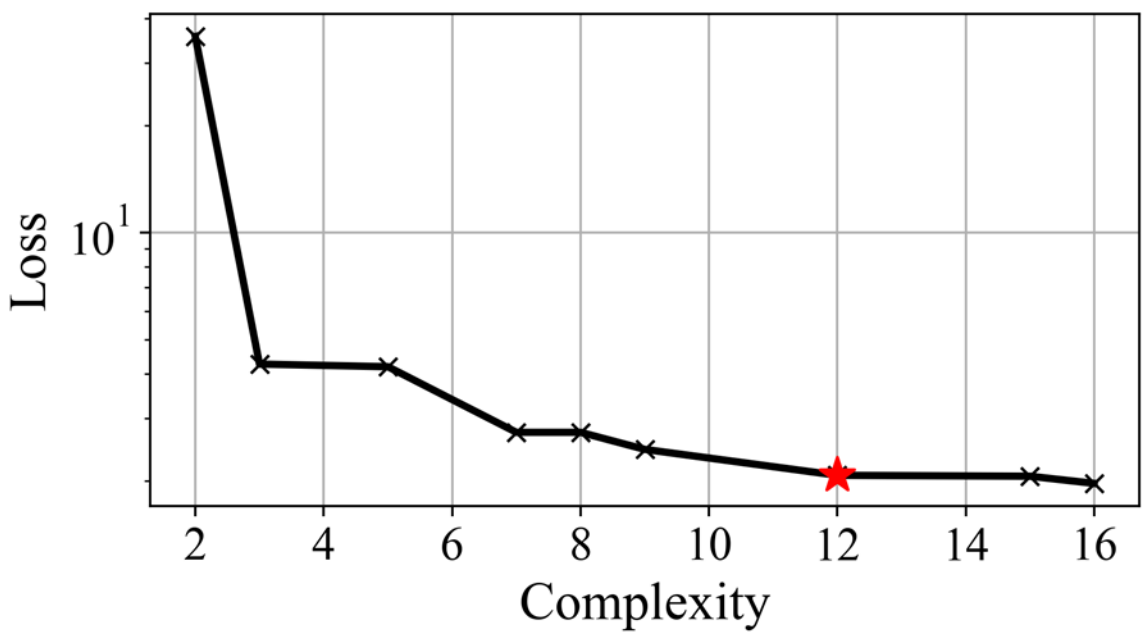

Figure 9. The loss-complexity plot of the final expressions

# 4. Test cases

In this section, the SST-CND3D model is tested in various 2D and 3D cases. In the 2D cases, the SST-CND3D model is expected to perform nearly identically to the SST-CND model, whereas in the 3D cases, the SST-CND3D model provides more accurate results than the SST-CND model.

## 4.1. 2D test cases

The SST-CND3D model is tested on the ZPG turbulent flat plate, the NASA hump case, and the NLR7301 multi-element airfoil [41]. The computational grid used for the ZPG turbulent flat plate is shown in Figure 10. It contains 133 faces along the wall and 97 cells in the wall-normal direction. The average $\Delta y^+$ of the first layer is approximately 0.2. The flow is incompressible, and OpenFOAM's SimpleFoam solver is used for this case. Figure 11 indicates that, in terms of the friction distribution along the plate and the velocity profile, the SST-CND3D model, the SST-CND model, and the SST model yield nearly identical outcomes. All three models produce skin friction consistent with the experimental data [42][43]. This result indicates that the SST-CND3D model retains the same accuracy as the baseline SST model in important wall-attached flows such as the ZPG turbulent flat plate.

The mesh used for the NASA hump case is shown in Figure 12. The flow is essentially incompressible, and OpenFOAM's SimpleFoam solver is used for this case. The Reynolds number based on the hump's chord length is approximately $9 \times 10^5$. A total of 1156 cells are used in the streamwise direction, and 304 cells are used in the wall-normal direction. The average $\Delta y^+$ of the first layer is approximately 0.2. Figure 13(a) shows that the SST-CND3D model performs nearly identically to the SST-CND model, outperforming the SST model in predicting the reattachment point. The profiles of the correction term $\beta$ are plotted in Figure 13(b). The correction term $\beta$ is defined as $\beta = (\beta_{CND} + \beta_{3D})f_d + 1$ for the SST-CND3D model and $\beta = \beta_{CND}f_d + 1$ for the SST-CND model. The almost identical profiles of $\beta$ indicate that $\beta_{3D}$ is nearly zero in 2D flows. Accordingly, the new $\beta_{3D}$ augmentation in the SST-CND3D model does not alter the already established performance of the SST-CND model in 2D incompressible flows.

The grid for the NLR7301 multi-element airfoil is shown in Figure 14. The Reynolds number based on the chord length of the clean airfoil is $\mathrm{Re} = 2.51 \times 10^6$, while the freestream Mach number is 0.185. The compressible solver CFL3D is used for this case. Figure 15 shows the $C_L - AOA$ curve computed by different models. The SST-CND3D and SST-CND models provide nearly identical $C_L$ predictions. Both models outperform the baseline SST model in predicting the stall of the multi-element airfoil. Figure 16 shows the $\beta$ distribution given by the SST-CND and SST-CND3D models. Both models activate the correction term in the mixing region above the flap, where the wake of the main wing and the jet formed by the slot are mixed. The nearly identical results of the SST-CND3D and SST-CND models in Figure 15 and Figure 16 demonstrate that the $\beta_{3D}$ augmentation does not affect the SST-CND model's established performance in 2D compressible flows.

Accordingly, these test cases demonstrate that the SST-CND3D model, with the deliberately constructed $\beta_{3D}$ correction term that vanishes in 2D scenarios, preserves the SST-CND model's high accuracy in important 2D wall-attached and complex separated flows. The sequential correction approach maintains the SST-CND model's performance in 2D cases.

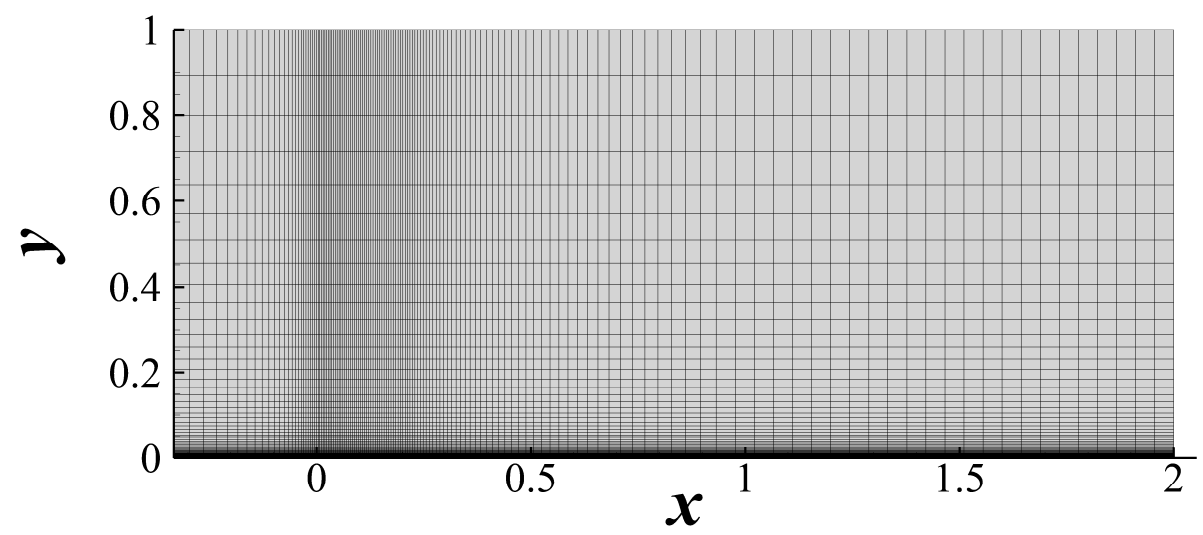

Figure 10. Computational grid of the ZPG turbulent flat plate case

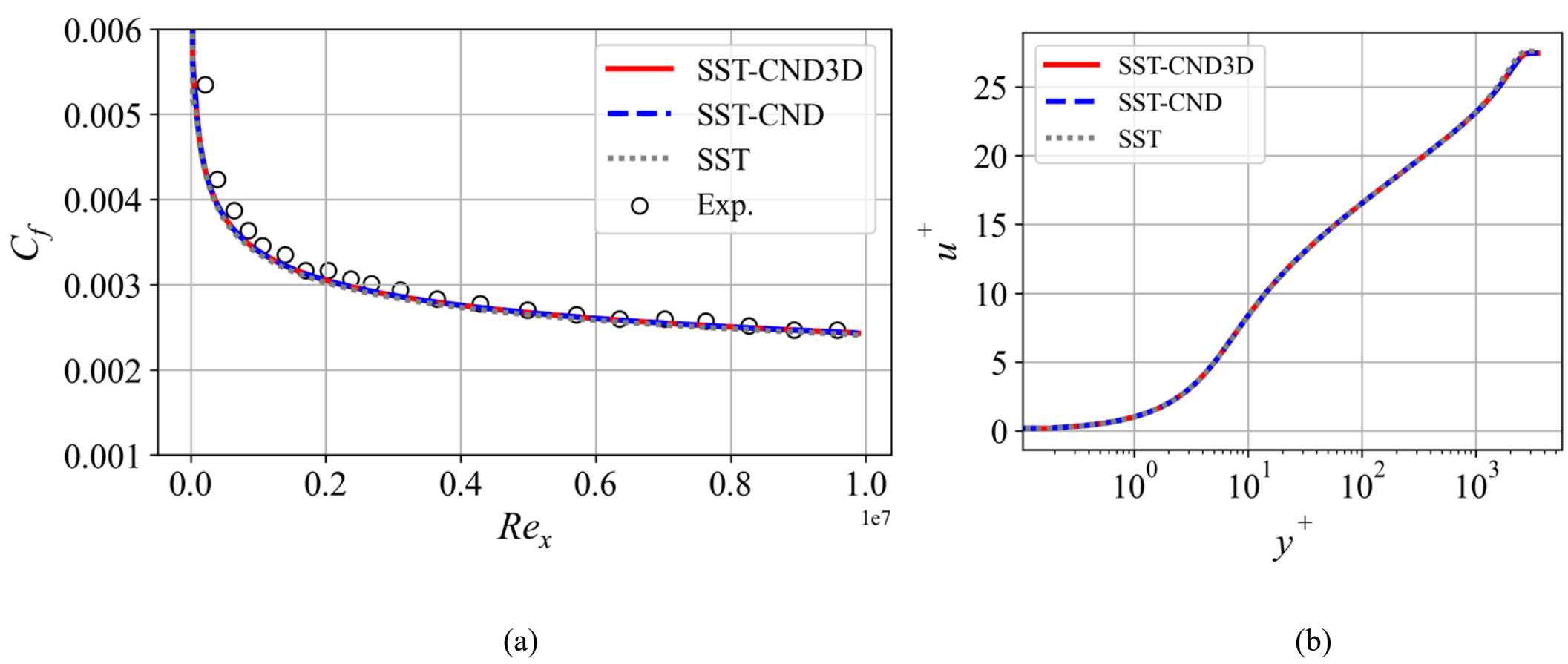

(a) (b)

Figure 11. Results of the turbulent flat plate, (a) $C_f$ distribution, (b) Velocity profile

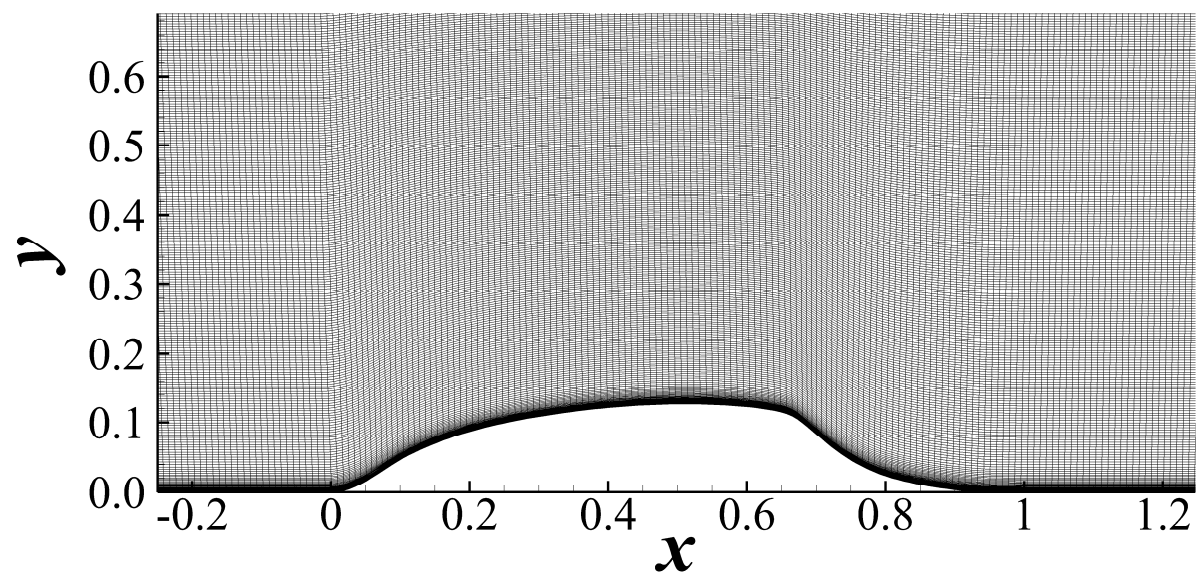

Figure 12. Computational grid of the NASA hump case

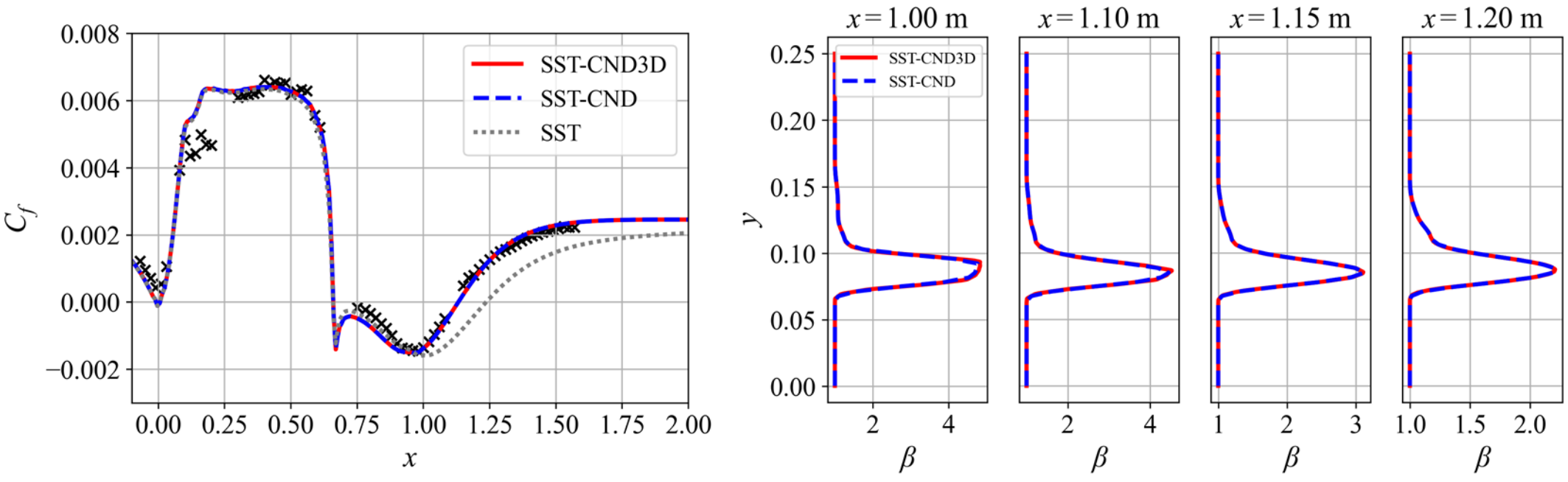

(a) (b)

Figure 13. Results of the NASA hump, (a) $C_f$ distribution, (b) $\beta$ profiles

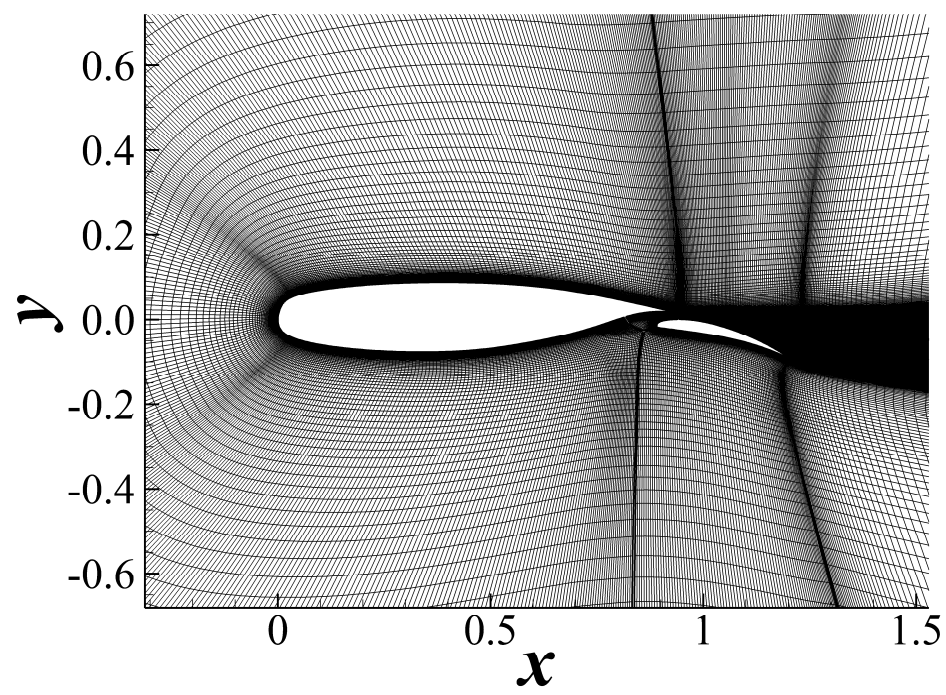

Figure 14. Computational grid of the NLR7301 multi-element airfoil

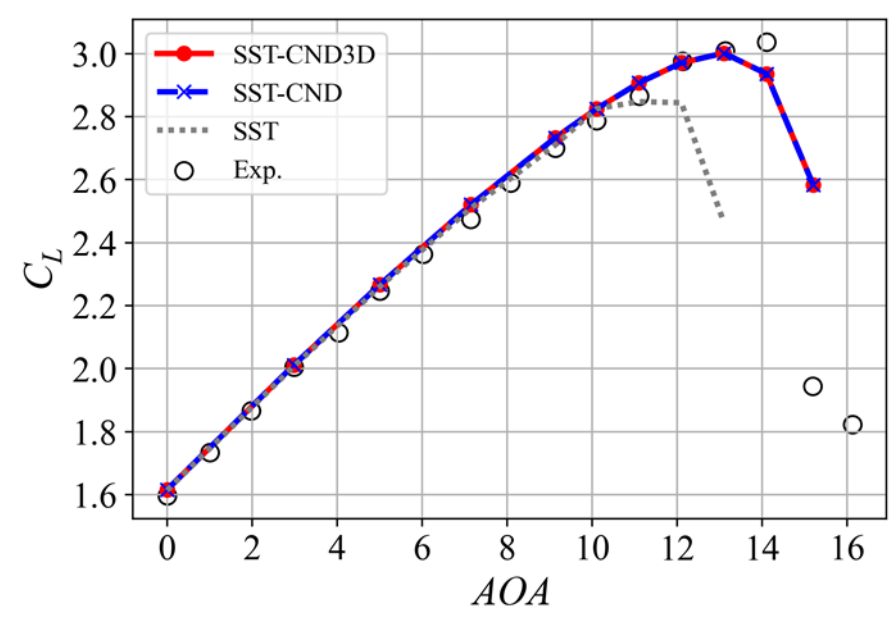

Figure 15. $C_L - AOA$ curve predicted by different models

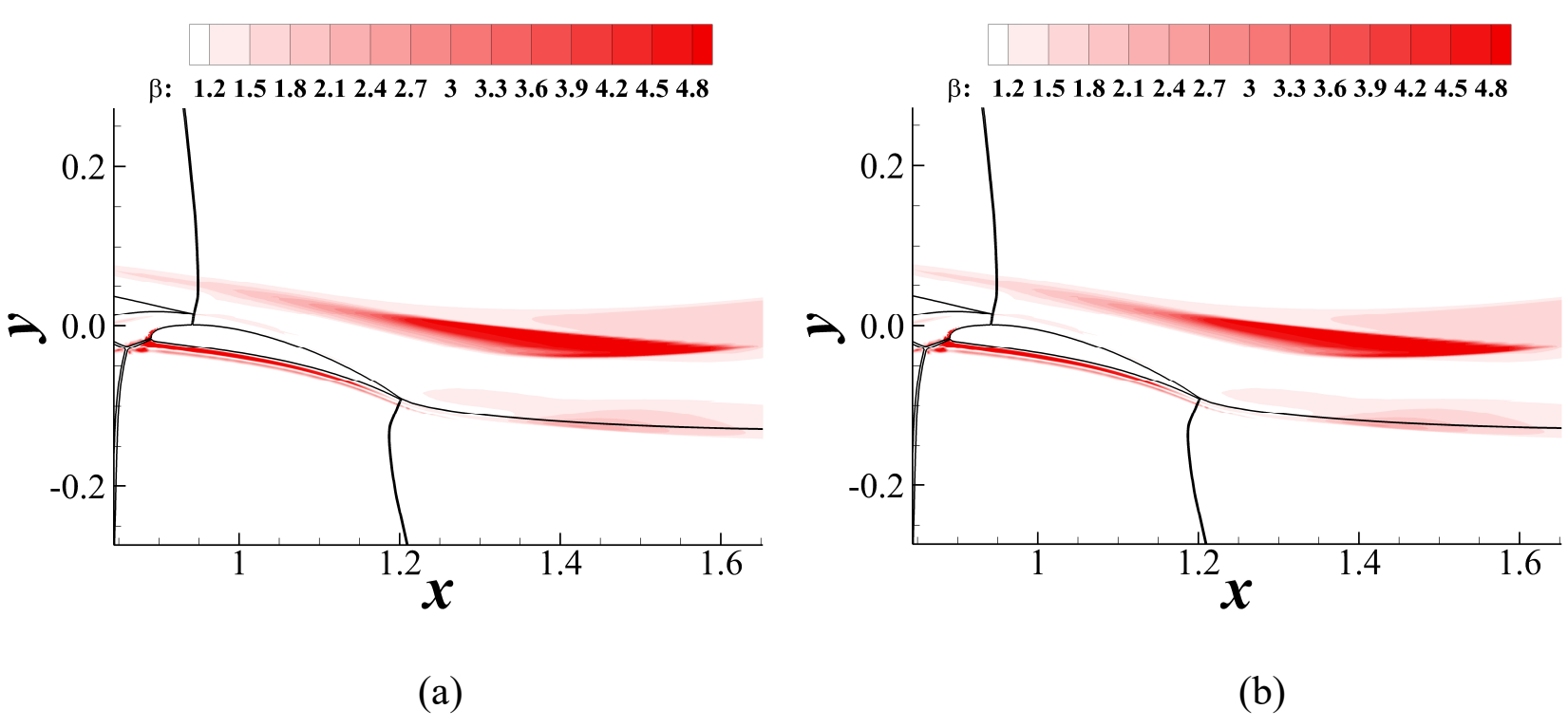

(a) (b)

Figure 16. $\beta$ contour given by (a) SST-CND model, (b) SST-CND3D model

## 4.2. Flow around a cube

The SST-CND3D model is applied to the training set of the $\beta_{3D}$ correction term, corresponding to the cube case. The computational grid used is identical to that employed in the field inversion process. The contours of $\beta$ obtained by the SST-CND model and the SST-CND3D model are illustrated in Figure 17 and Figure 18, respectively. Unlike the 2D cases, the SST-CND3D model produces a considerably larger $\beta$ compared to the SST-CND model in the separated shear layer around the cube. A more distinct $\beta$ is also observed in front of the cube. The reattachment point on the symmetry plane predicted by the SST-CND3D model moves forward compared to the prediction of the SST-CND model. The difference between Figure 17 and Figure 18 indicates that the $\beta_{3D}$ term is activated in 3D scenarios. The velocity profiles and the $C_p$ distribution predicted by the SST-CND3D model are closer to the experimental data than those predicted by the SST-CND model, as shown in Figure 19(a) and (b).

Accordingly, the results demonstrate that the SST-CND3D model achieves higher accuracy compared to the SST-CND model in the 3D training set of the $\beta_{3D}$ augmentation term. The augmentation term $\beta_{3D}$ generated by the sequential correction approach effectively improves the SST-CND model's performance in 3D scenarios.

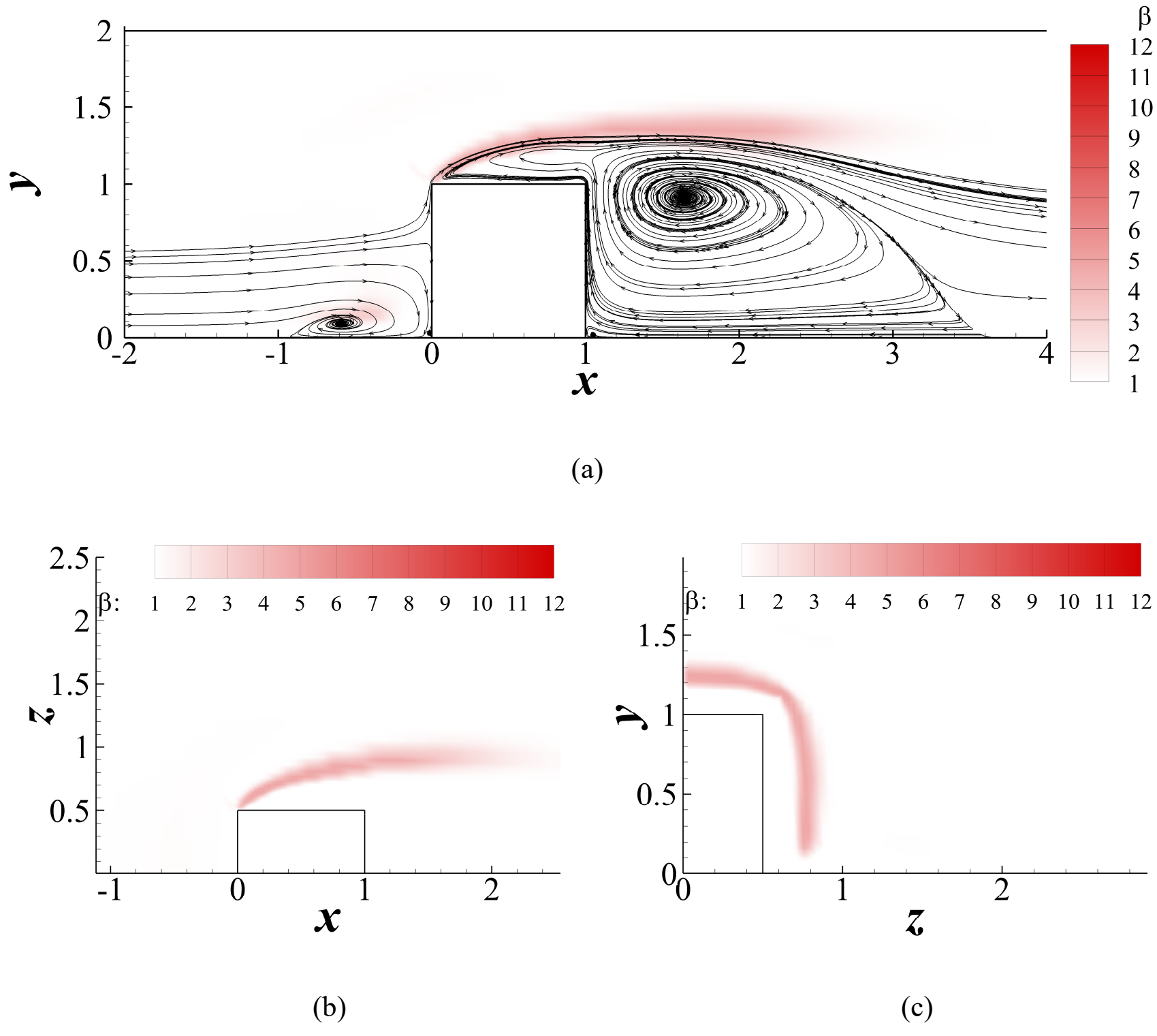

Figure 17. Contour of $\beta$ obtained by the SST-CND model on the: (a) Symmetry plane ($z = 0$), (b) $y = 0.5\ m$ plane, (c) $x = 0.5\ m$ plane.

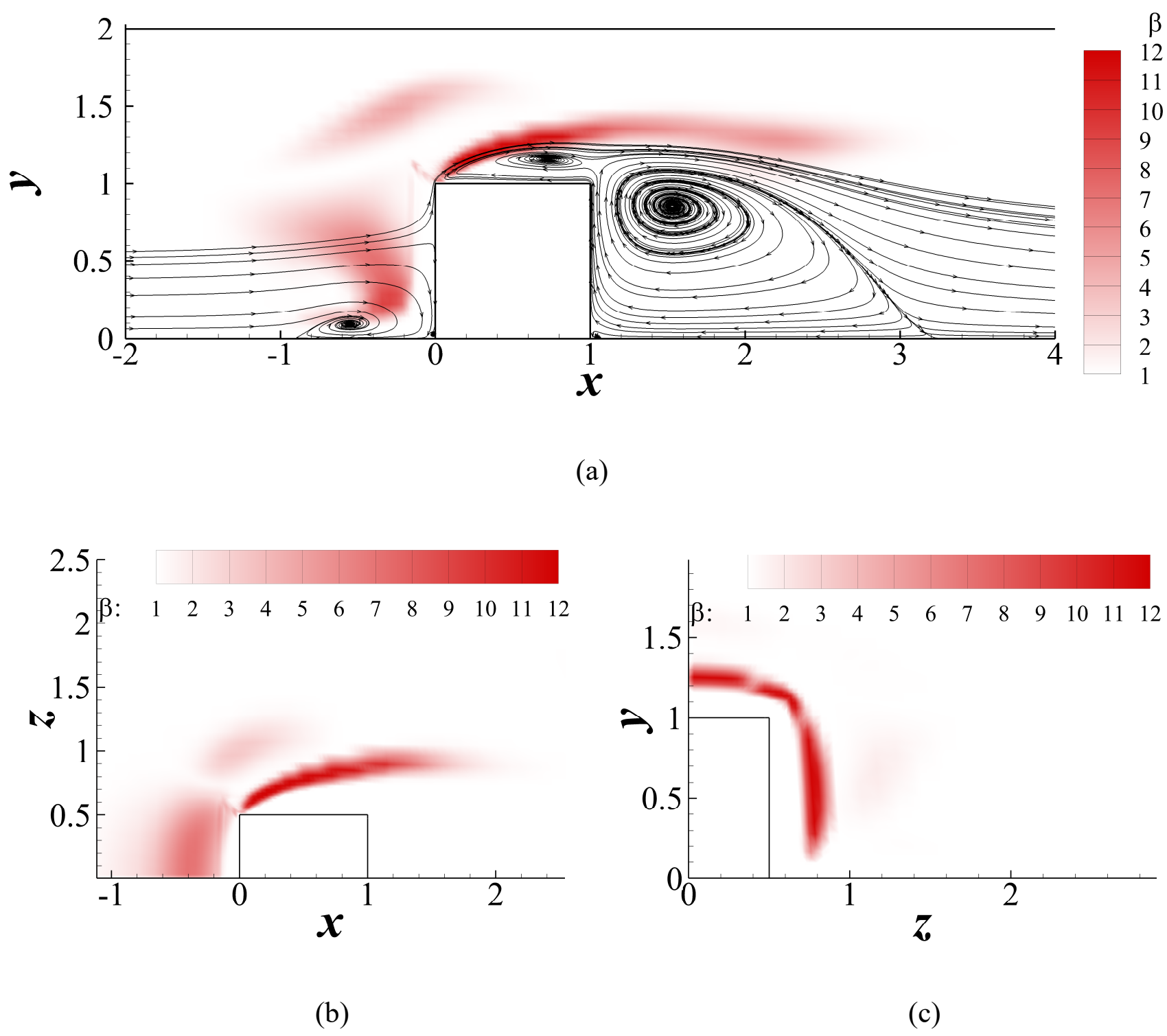

Figure 18. Contour of $\beta$ obtained by the SST-CND3D model on the: (a) Symmetry plane ($z = 0$), (b) $y = 0.5\ m$ plane, (c) $x = 0.5\ m$ plane.

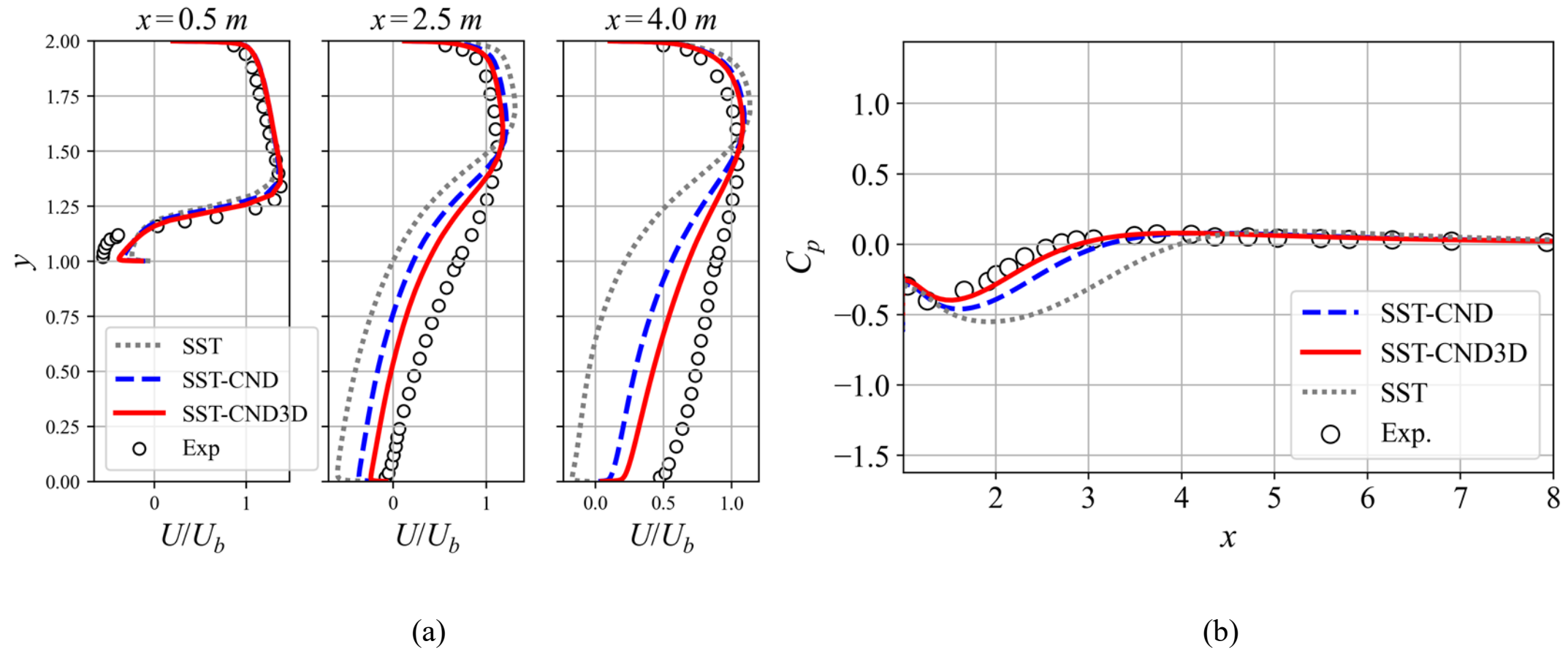

Figure 19. Results of different models, (a) Velocity profiles, (b) $C_p$ distribution

## 4.3. Flow around an axisymmetric hill

In this section, the SST-CND3D model is applied to an axisymmetric hill whose experimental data were obtained by Simpson et al. [44]. The definition of the geometry of the hill can be found in Ref. [44]. The height of the hill is $H = 0.078\ m$. The Reynolds number based on $H$ and a reference speed (the nominal freestream speed of the wind tunnel) of $U_{ref} = 27.5\ m/s$ is $1.3 \times 10^5$. In the absence of the hill, a zero-pressure gradient boundary layer with thickness $\delta = H/2$ is presented at the location of the hill. LES simulations were also carried out for this case by Garcia-Villalba, M. et al. [45]. Here, we use a similar computational domain as in Ref. [45], except that we only compute half of the domain since the flow is symmetric. Our computational domain is $5.85H$ wide, $3.2H$ high, and $20H$ in the streamwise direction. The hill is placed $4H$ downstream of the inlet. The computational domain is shown in Figure 20. The total number of cells is around $1.3 \times 10^6$. The average $\Delta y^+$ is around 0.25. At the inlet, the velocity profile, the turbulent kinetic energy profile, and $\omega$'s profile are specified to make sure the boundary layer thickness $4H$ downstream of the inlet in the absence of the hill is approximately $H/2$.

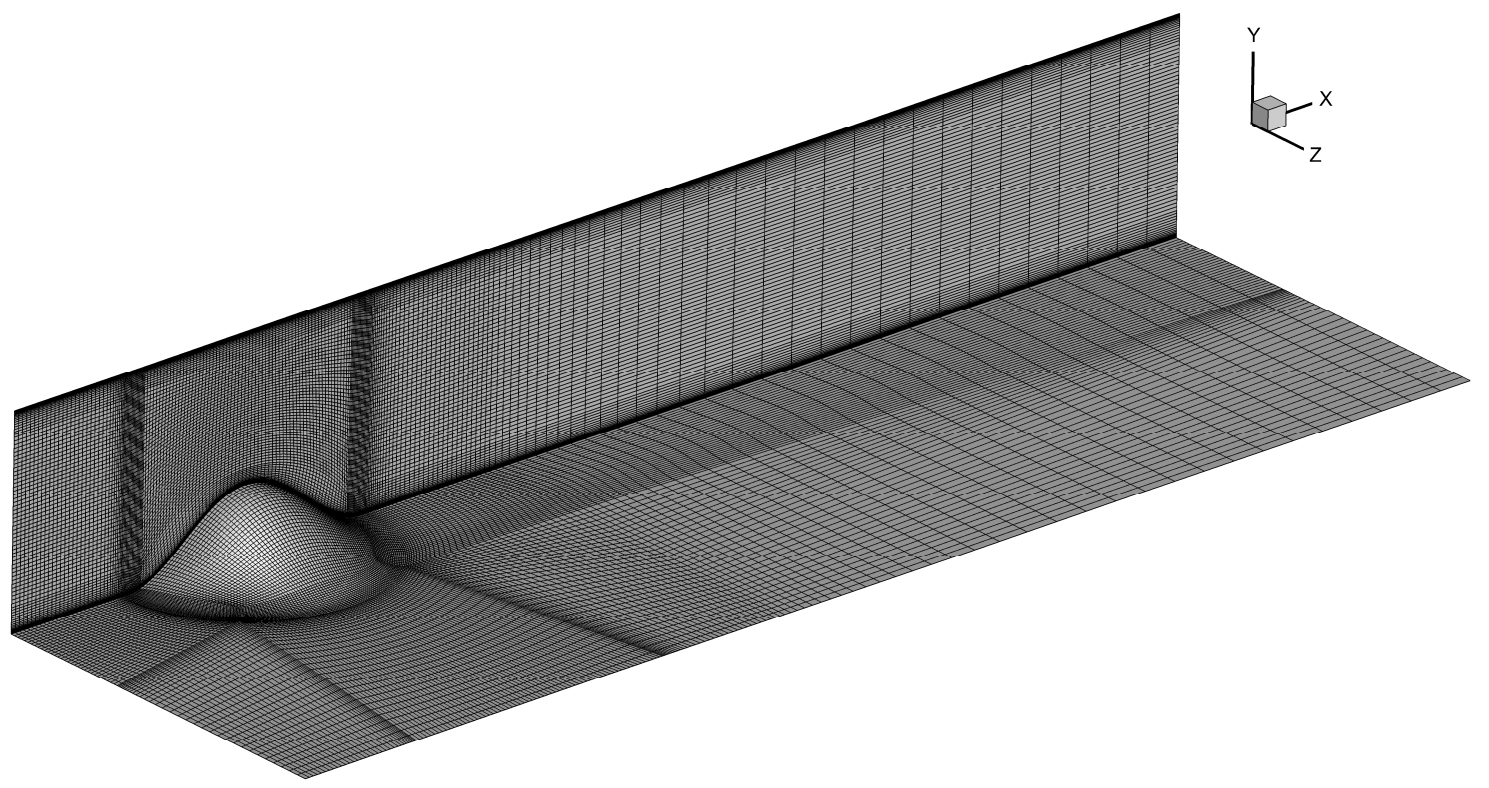

Figure 20. The computational mesh of the axisymmetric hill

The velocity profiles at different spanwise locations at the same streamwise location ($x = 3.63H$) are shown in Figure 21. The SST-CND and the SST-CND3D models both outperform the SST model. Compared to the SST-CND model, the SST-CND3D model predicts a velocity profile that is closer to the experimental data. The pressure coefficient along the centerline of the hill is shown in Figure 22. Both the SST-CND and the SST-CND3D models show improvement over the prediction given by the SST model, with the SST-CND3D model obtaining a $C_p$ distribution that is slightly closer to the experiment. The surface streamline plots are compared to the experiment results [46] in Figure 23. It shows that all three turbulence models predict an early separation line, while the SST model gives the largest recirculation zone. The comparison of the separation line is shown in Figure 24. It confirms that there's a minor gain produced by the SST-CND3D model compared to the SST-CND model. The contours of the correction term $\beta$ are shown in Figure 25. It can be seen that the SST-CND3D model gives a much more pronounced activation of $\beta$ in the separated shear layer behind the hill.

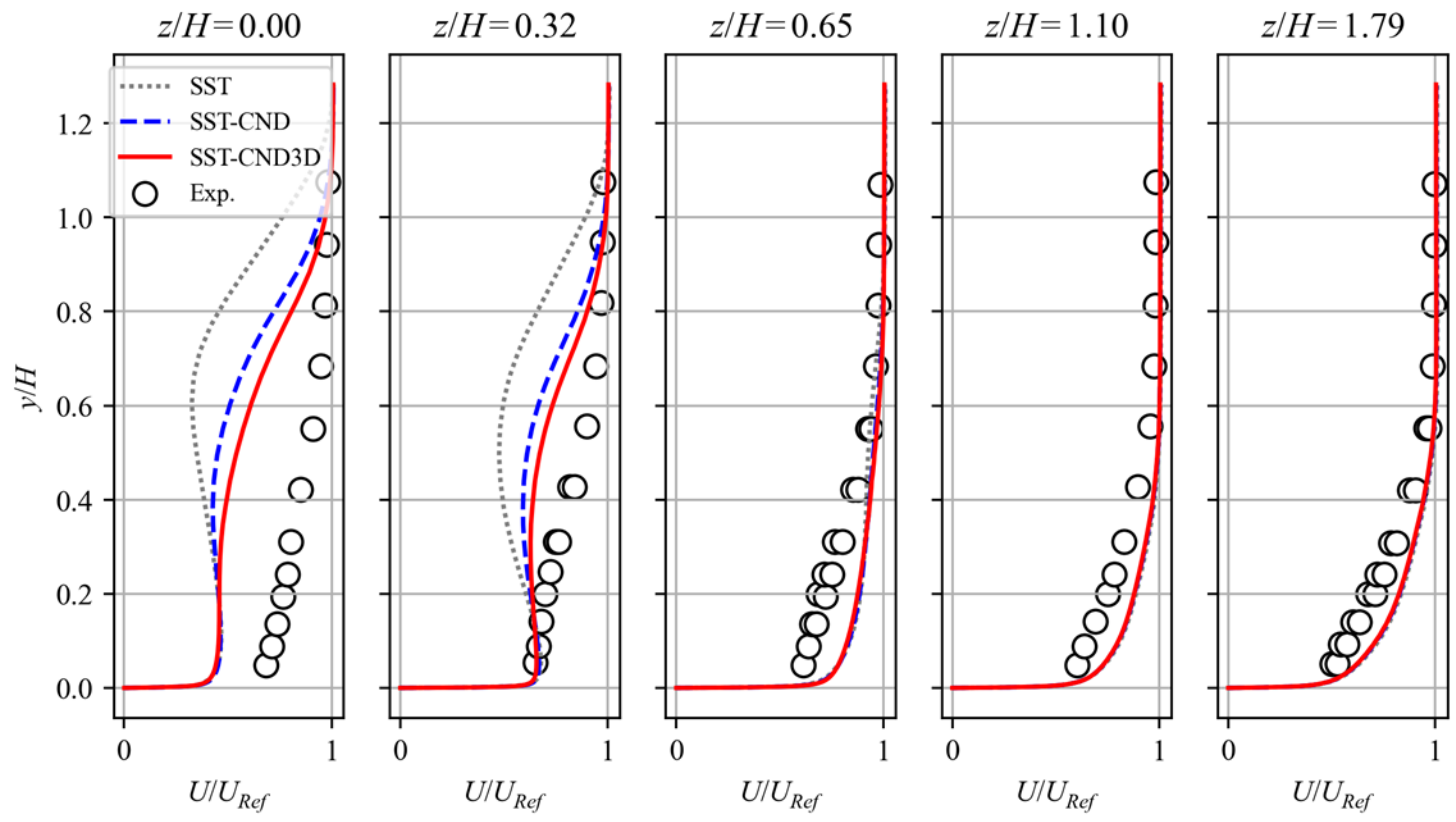

Figure 21. The velocity profiles at $\frac{x}{H} = 3.63$, with different spanwise locations

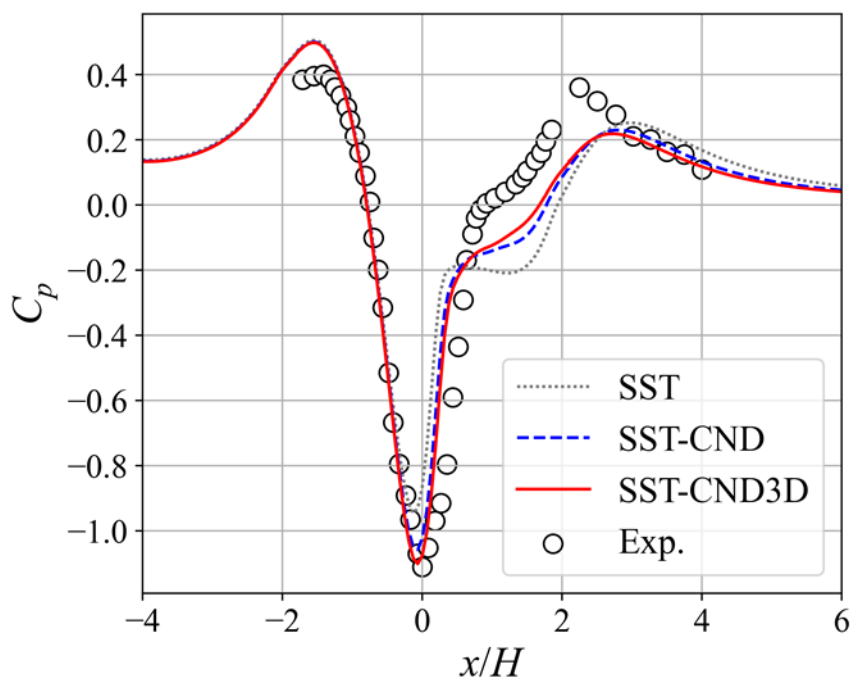

Figure 22. The $C_p$ distribution along the streamwise direction at the centerline

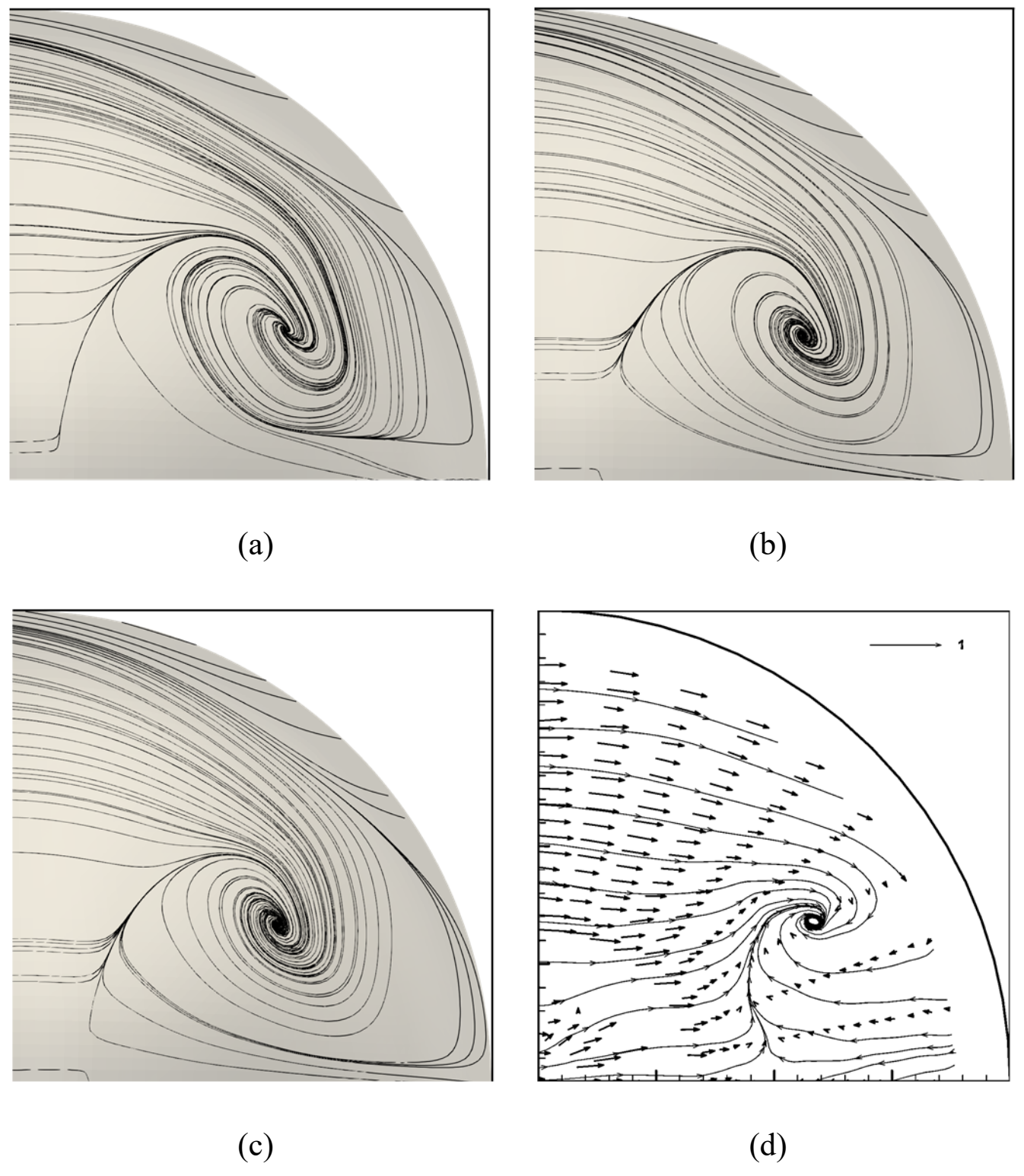

Figure 23. The surface streamline plots given by (a) the SST model, (b) the SST-CND model, (c) the SST-CND3D model, and (d) the experiment (taken from Ref. [46]).

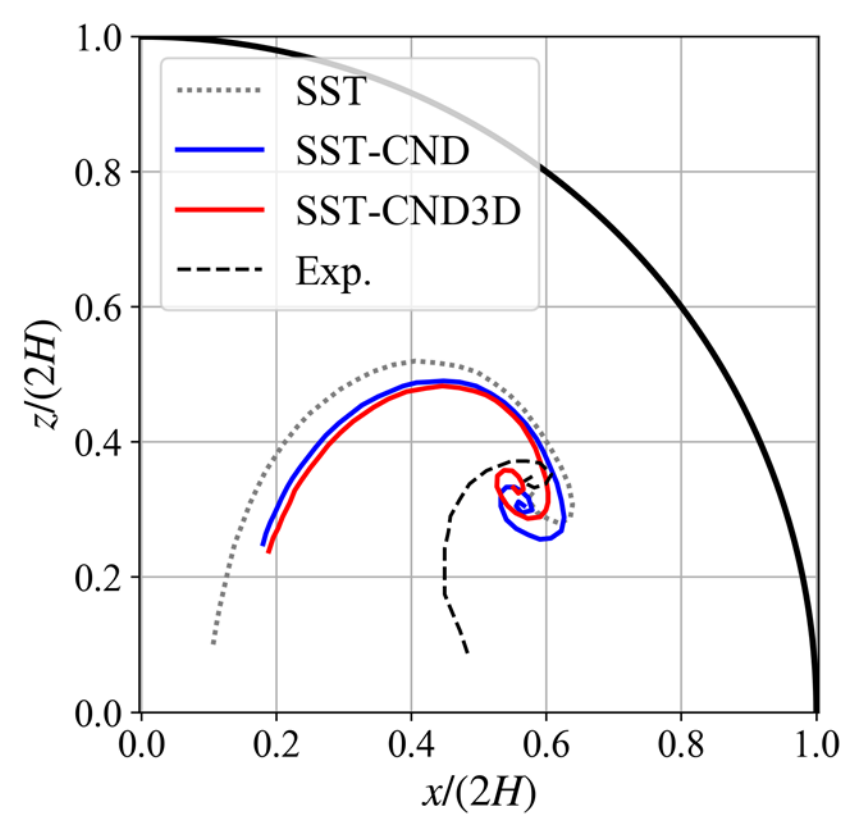

Figure 24. The overlain separation line of different models for comparison

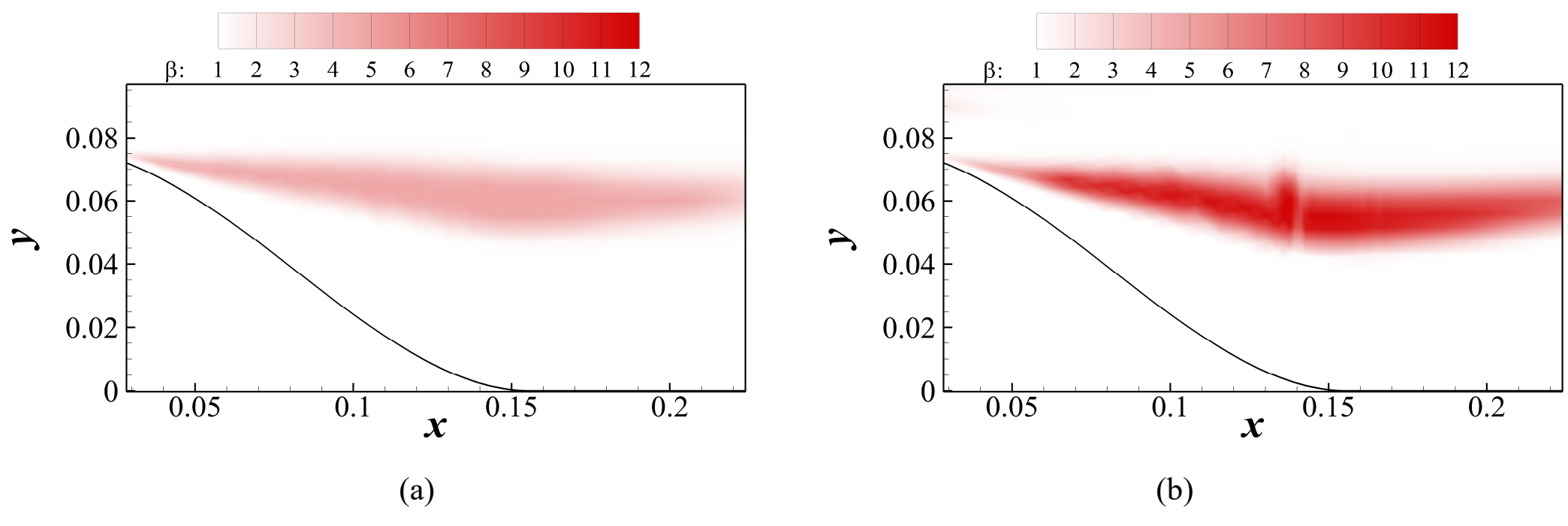

Figure 25. The contour of the correction term $\beta$ given by (a) the SST-CND model, and (b) the SST-CND3D model

In summary, the SST-CND and the SST-CND3D models both give results that are much better than the SST model in the axisymmetric hill case, with the SST-CND3D model showing improvement over the SST-CND model in this 3D case.

## 4.4. Flow around the NASA FAITH hill

We apply the SST-CND3D model to the NASA FAITH (Fundamental Aero Investigates The Hill) hill [21] case in this section. The definition of the geometry can be found in Ref.[21]. Compared to the previously investigated axisymmetric hill case, the FAITH hill features a taller and thinner geometry, with a larger Reynolds number. The Reynolds number based on the hill height ($0.1524\ m$) and the freestream velocity ($50.292\ m/s$) is $5 \times 10^5$. The computational domain and mesh are shown in Figure 26. About 3.3 million cells are used. The inlet velocity profiles and the turbulent quantities profiles are given so that the boundary layer thickness at the inlet is approximately 1/3 of the hill height. The no-slip boundary condition is imposed on both the upper wall and the lower wall. For the side boundary, the symmetry boundary condition is imposed. Only half of the model is computed, considering the symmetric property of the flow.

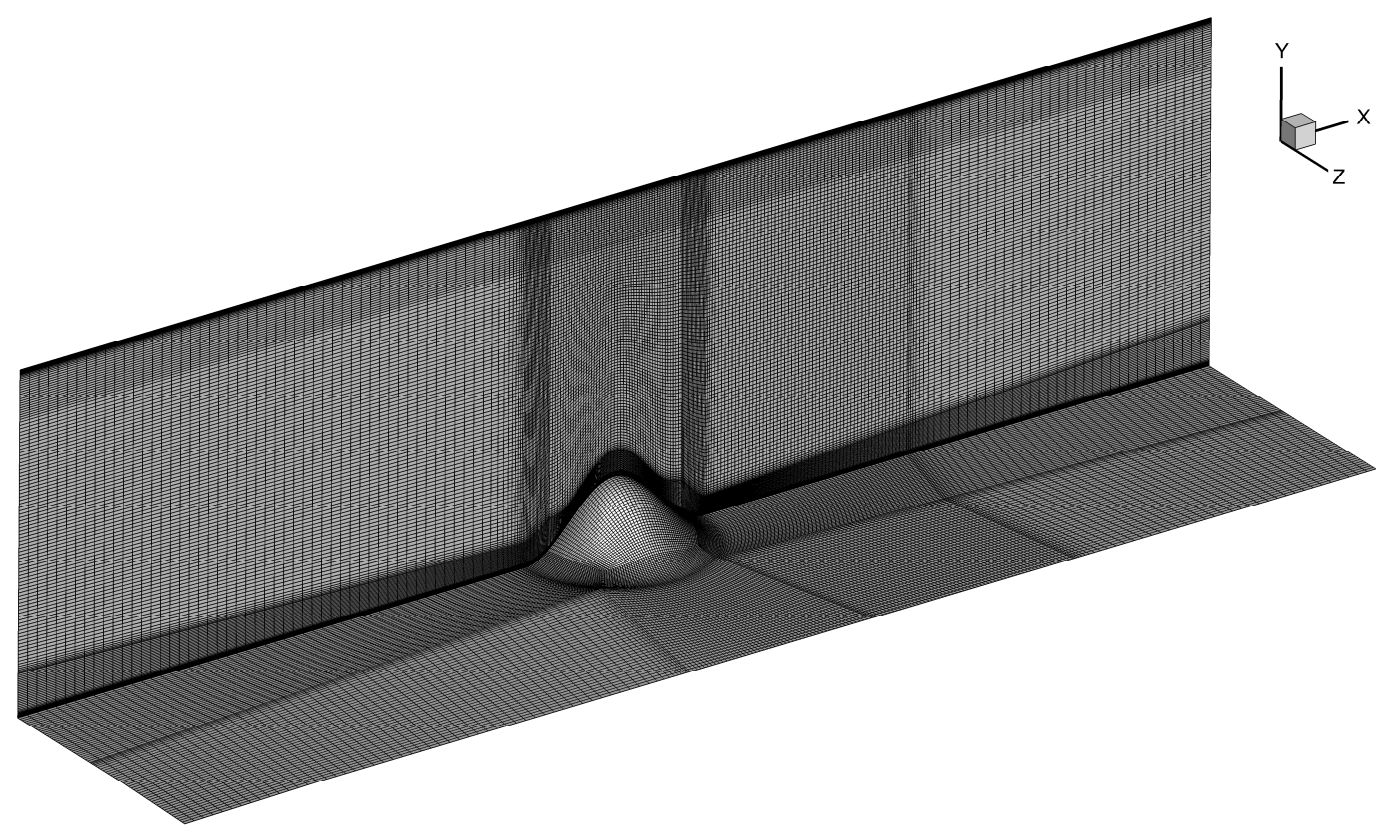

Figure 26. The computational mesh of the NASA FAITH hill

The velocity profiles on the symmetry plane are shown in Figure 27. It can be seen that both the SST-CND3D and the SST-CND models give a better prediction compared to the SST model, with the SST-CND3D model even closer to the PIV experiment data. This indicates that the SST-CND3D model indeed provides additional correction capability compared to the baseline SST-CND model, showing better generalizability to 3D flows. Figure 28 plots the $x$-direction (streamwise) friction coefficient ($C_{fx}$) distribution along the centerline of the hill. While all three models give separations that are too early, it can be seen that the SST-CND3D model predicts a separation point that is closest to the experimental data [47]. The surface vector plots given by the experiment [47] and different models are shown in Figure 29. The SST model predicts a recirculation zone that is quite thin and has a large aspect ratio. The recirculation zones given by the SST-CND and the SST-CND3D model are smaller and wider, which are somewhat closer to the experiment's small recirculation zone. Figure 30 is the contour of the correction term $\beta$ on the symmetry plane. Again, the correction regions of the SST-CND3D and the SST-CND model are essentially the same, but the SST-CND3D model features a more pronounced increase of the correction term $\beta$ in the separated shear layer.

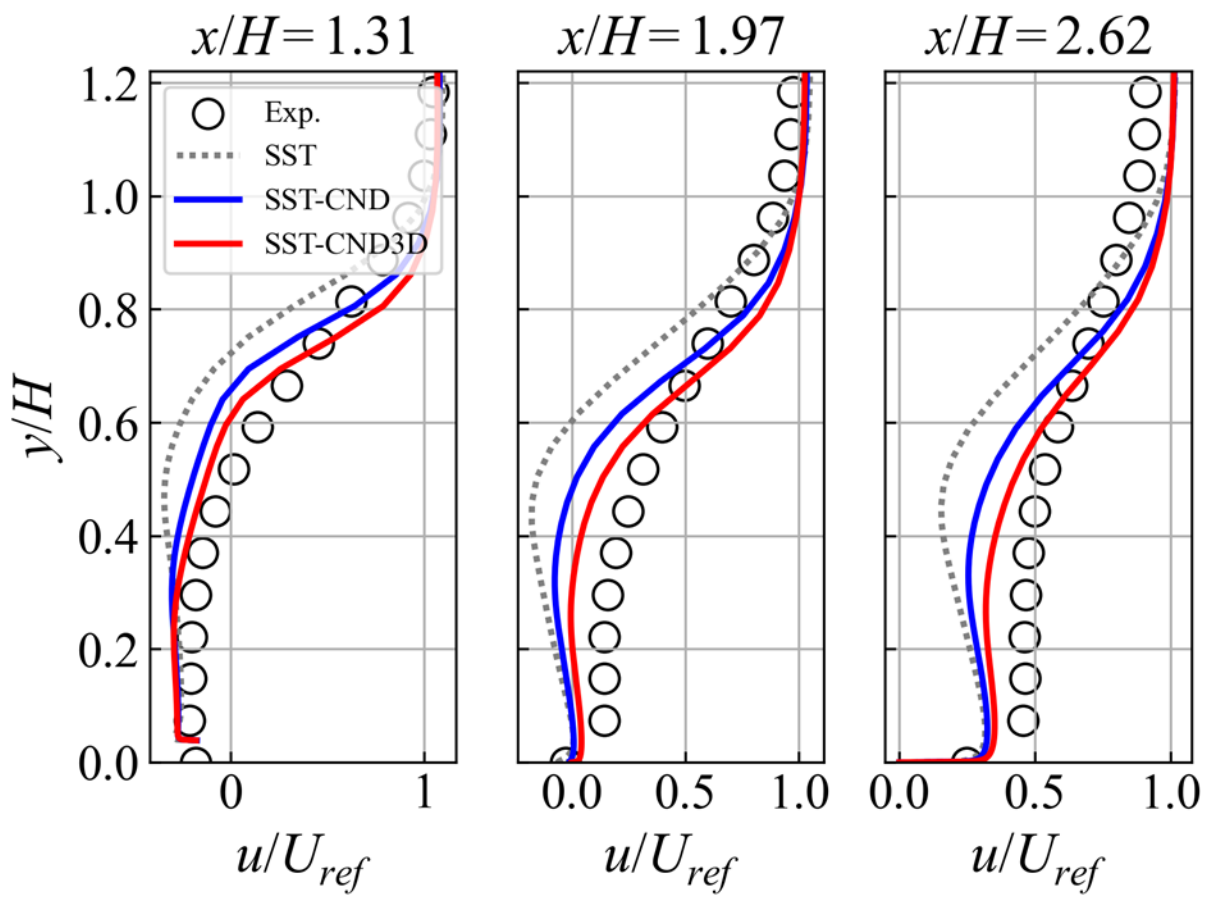

Figure 27. The velocity profiles on the symmetry plane

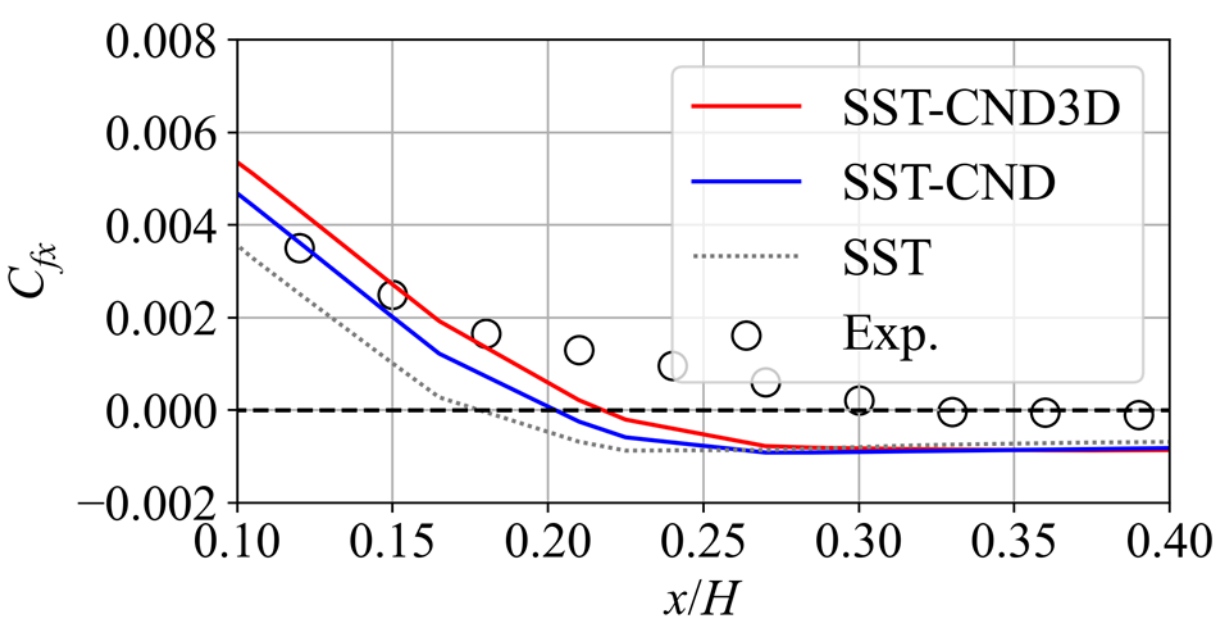

Figure 28. The friction coefficient distribution along the centerline near the separated region

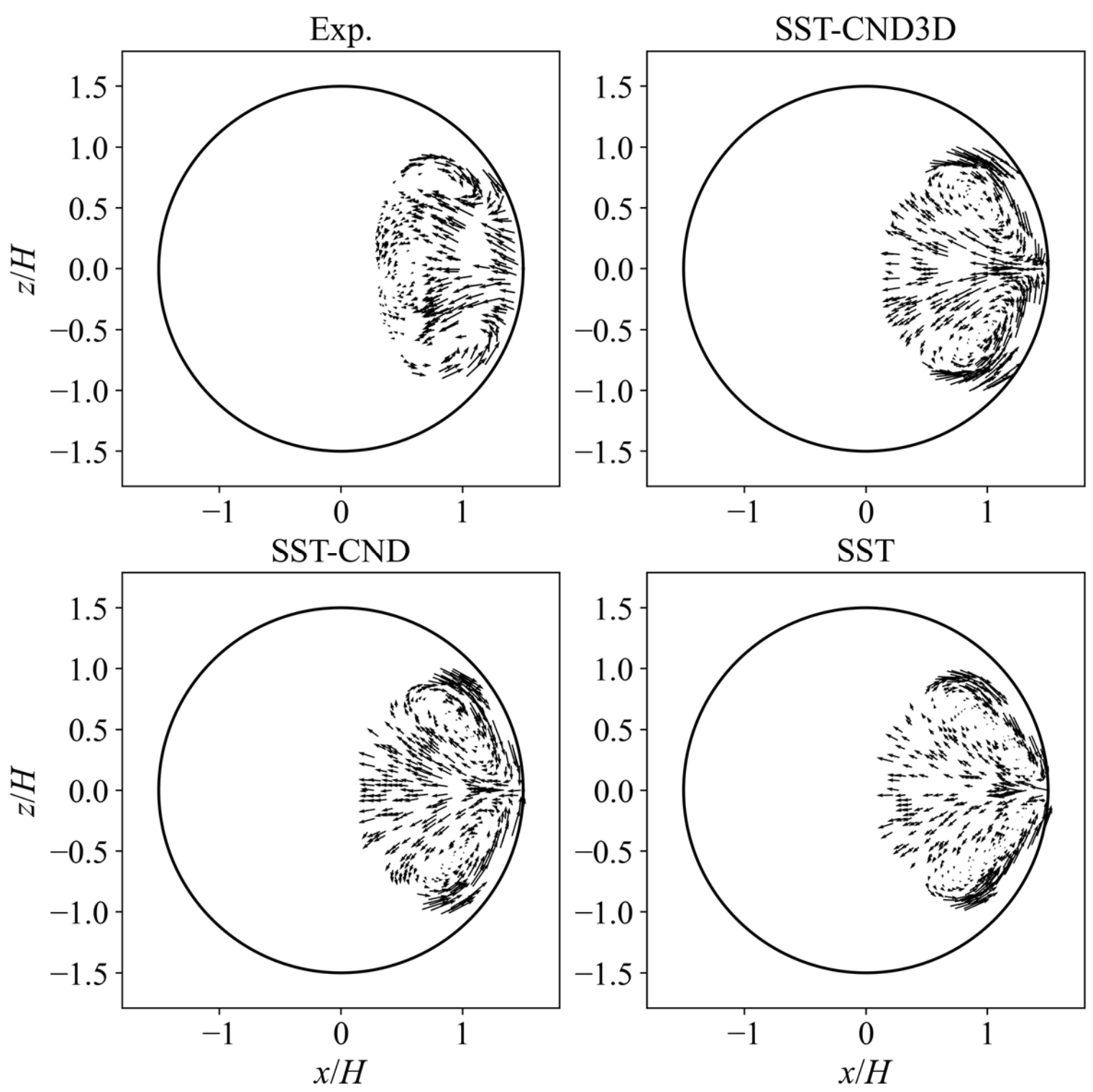

Figure 29. The surface vector plot of the hill in the separation zone, the experimental data is from Ref. [47]

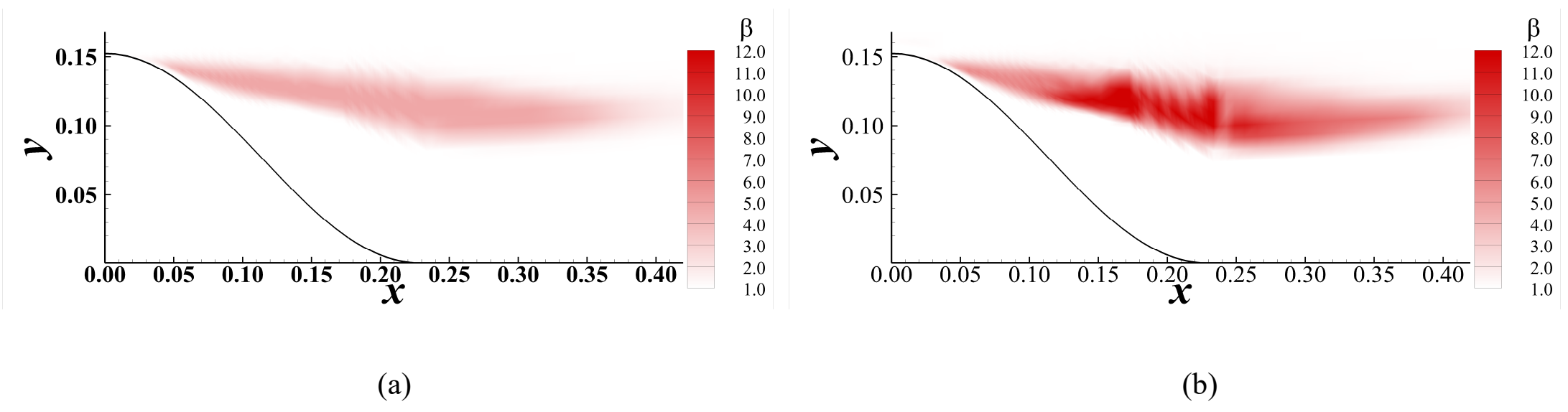

Figure 30. The correction term $\beta$ given by the (a) SST-CND model, (b) SST-CND3D model

In general, the SST-CND3D model performs better than the SST-CND model in the NASA FAITH hill case, showing the effectiveness of the proposed 3D correction.

## 4.5. Flow around a complex 3D high-lift configuration

In this section, the model is tested using the JAXA Standard Model (JSM) [48], which represents a complex 3D high-lift configuration. The geometry of the JSM is depicted in Figure 31. A 3D structured grid is utilized for computation, as illustrated in Figure 32. The nacelles and pylons are included in the model, as well as the flap track and slat track fairings. A total of 96 million grid cells is used, and the average $\Delta y^+$ of the first layer is approximately 0.4. The average grid spacing in the streamwise direction is approximately 0.75% $c_{ref}$, where $c_{ref}$ is the reference chord length and equals 0.5292 m [48]. The average grid spacing in the spanwise direction is approximately 1% $c_{ref}$. The freestream Mach number is 0.172, and the Reynolds number based on the mean aerodynamic chord is 1.93 million.

Initially, the force coefficients predicted by different models are compared. Figure 33(a) exhibits that the $C_L$ at large $AOA$ predicted by the SST-CND3D model is closer to the experimental data than the $C_L$ predicted by the SST-CND model. Table 4 confirms that the SST-CND3D model provides more accurate predictions of $C_{L,max}$ and $AOA_{stall}$. The error of $C_{L,max}$ decreases by nearly 50% compared to the SST-CND model. The polar curve in Figure 33(b) also indicates a better prediction of $C_D$ at large $C_L$. In addition, the SST-CND3D model provides a more accurate $C_m$ compared to the SST-CND model at high $C_L$, as shown in Figure 33(c). This finding indicates that the SST-CND3D model achieves a more realistic force distribution along the JSM body.

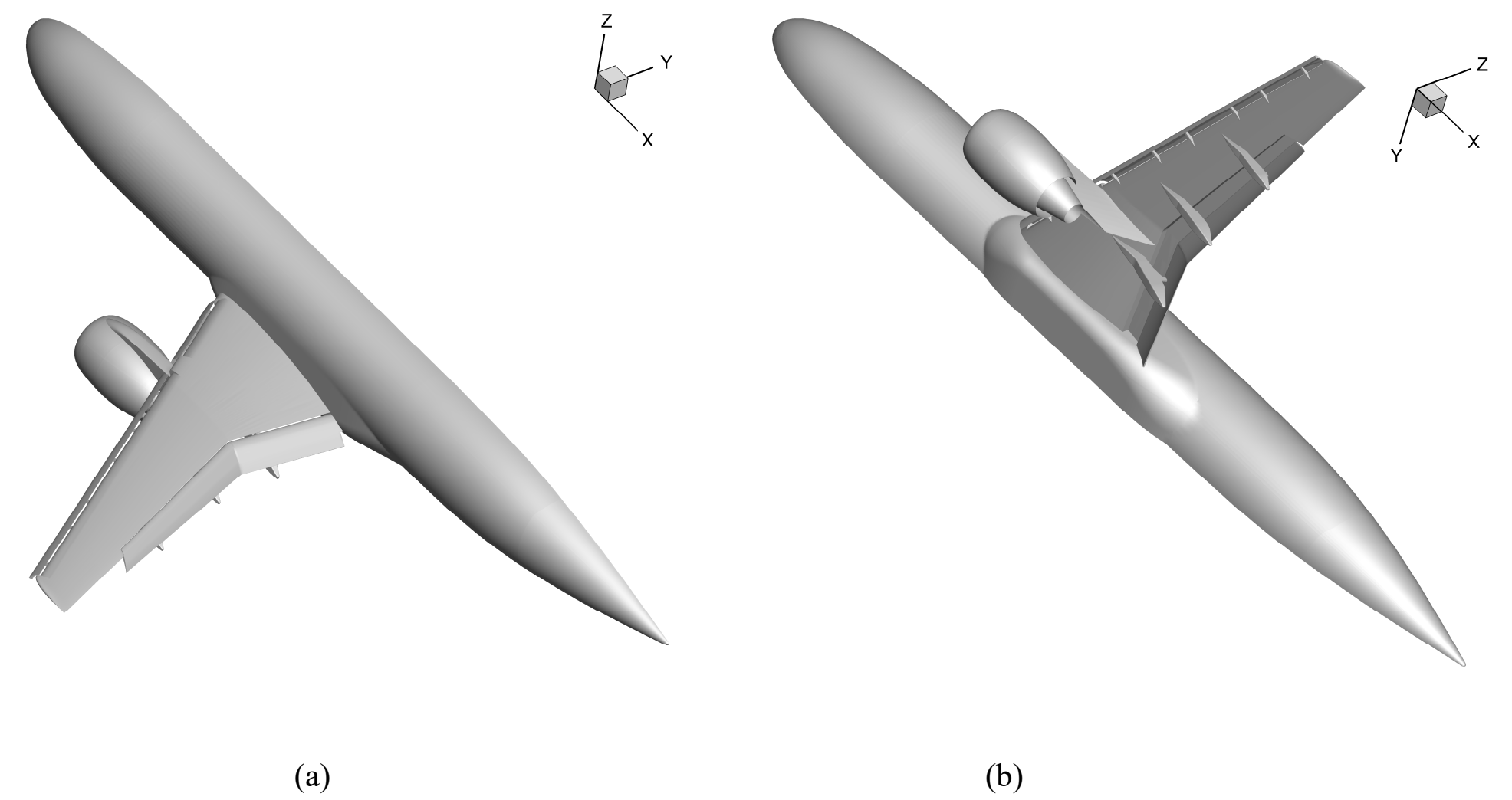

(a) (b)

Figure 31. Geometry of the JAXA standard model (JSM), (a) Upper view, (b) Lower view

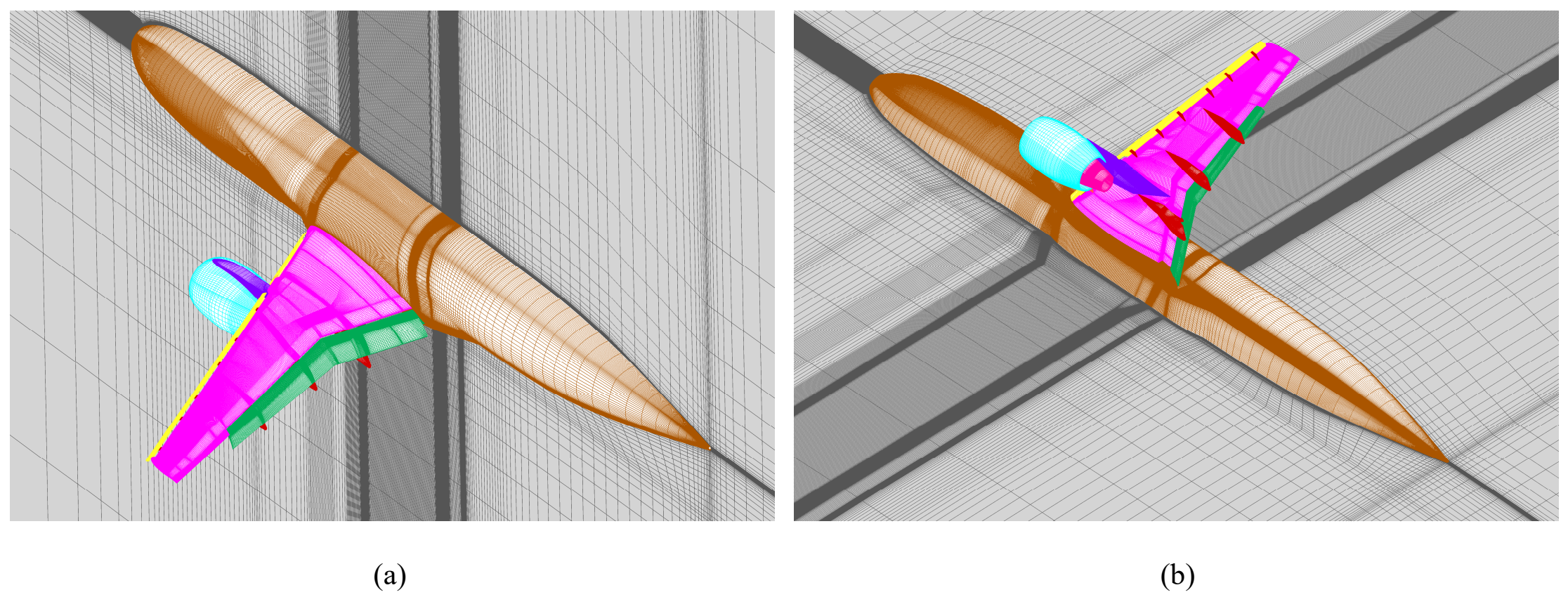

(a) (b)

Figure 32. 3D structured grid used for computation, the grid is coarsened for clarity, (a) Upper view, (b) Lower view

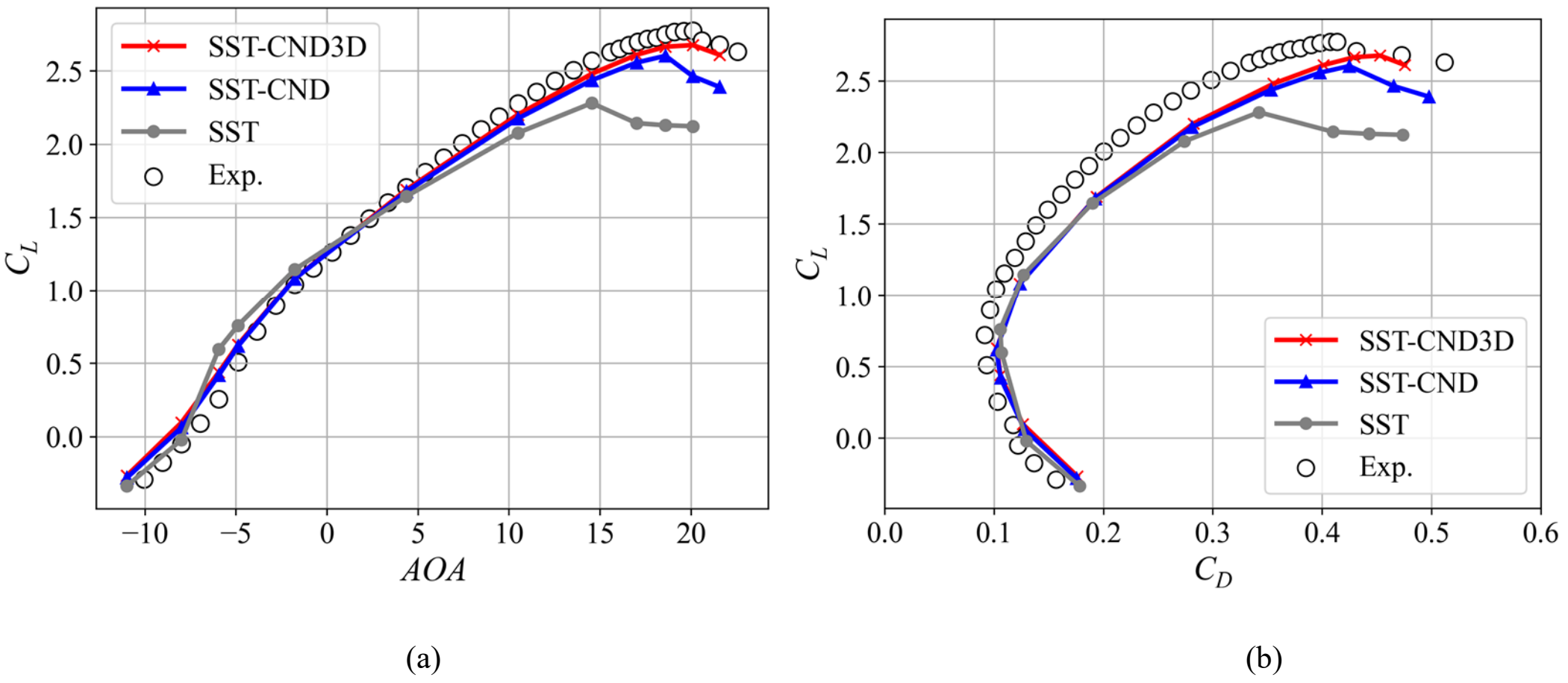

(a) (b)

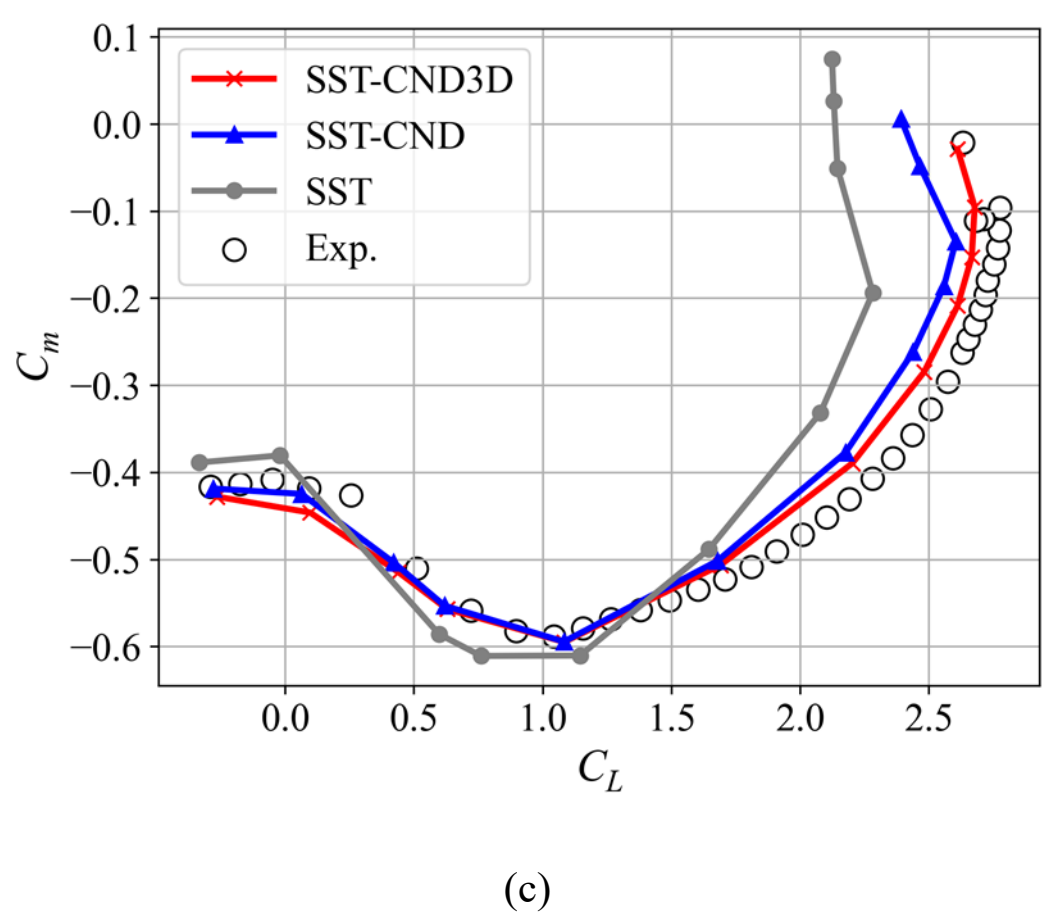

(c)

Figure 33. Force coefficients predicted by different models, (a) $C_L - AOA$ curve, (b) $C_L - C_D$ curve, (c) $C_m - C_L$ curve

Table 4. Prediction value and error of the maximum lift coefficient and stall angle of attack

| | SST | SST-CND | SST-CND3D | Exp. |
|---|---|---|---|---|
| $AOA_{stall}$/difference | 14.54°/5.55° | 18.59°/1.5° | 20.09°/0° | 20.09°/- |
| $C_{L,max}$/error | 2.282/17.8% | 2.604/6.2% | 2.677/3.5% | 2.775/- |

Figure 34 presents the surface streamline plot of the JSM at $AOA = 18.6°$. It can be observed that the SST model predicts a large separation zone in the middle of the wing. The SST-CND and SST-CND3D models both predict attached flow, consistent with the oil flow image in Figure 34(d). The surface streamline plots of the SST-CND and SST-CND3D models appear similar; however, the SST-CND3D model predicts a relatively smaller separation in the outboard region of the wing and at the wing root, where the 3D effect is more pronounced. Figure 35 shows the pressure coefficient distribution along six different sections of the wing at $AOA = 18.6°$. The SST model underpredicts the suction peak of the wing, and a flat $C_p$ distribution caused by flow separation is visible at section E–E. Both the SST-CND and SST-CND3D models yield results that agree better with the experimental data. However, at section H–H, all models underpredict the suction peak, with the SST-CND3D model performing slightly better. Figure 36 displays the $C_p$ distribution at $AOA = 20.6°$, corresponding to the post-stall condition. At section E–E (the middle of the wing), the SST and SST-CND models both produce a flat $C_p$ distribution resulting from flow separation, while the SST-CND3D model performs significantly better and matches the experimental data. At the outboard sections of the wing (G–G and H–H), the SST-CND3D model also provides a stronger suction that corresponds more closely with the experiment. We compare the $\beta$ contours at the leading edge of the wing and the slat at the E-E section (Figure 37) and H-H section (Figure 38). At the E-E section, the SST-CND3D model gives a substantially higher suction compared to the SST-CND model. This is also reflected in the $\beta$ contours in Figure 37, where the SST-CND model gives a separated shear layer (activated $\beta$) at the middle of the chord, but the SST-CND3D model does not. At the H-H section, in Figure 38, both models give a separated shear layer at the middle of the chord, which corresponds to the underestimated suction at H-H. The $C_p$ distribution in Figure 36 indicates that the correction term is more effective at the E-E section (midboard) than at the H-H section (outboard). This is because the separation in the midboard region is caused by the slat track shown in Figure 39. In this region, the flow is hindered by a sharp object, and the spanwise pressure gradient is not strong, which resembles the cube training case. This makes the model effective in the midboard region. In the outboard region, on the other hand, the influence of spanwise pressure gradient and the wing-tip vortex becomes important. The flow field no longer resembles the training case, making the model less effective. The surface streamline plots in Figure 40 confirm that the flow topologies at the middle of the wing given by different models are significantly different. While the SST-CND3D model only predicts a small triangular recirculation zone, the SST-CND model gives a relatively large recirculation at the mid-wing. The recirculation of the SST model is even larger. This difference is reflected in the pressure distribution of the E-E section, which resides in the recirculation zone.

Based on the analysis above, we can say that the SST-CND3D and the SST-CND models tend to predict smaller recirculation than the SST model. The recirculation zone given by the SST-CND3D model is even smaller. This is because an increased $\beta$ can cause larger destruction of $\omega$, which in turn increases $k$ and reduces $\omega$. $\nu_T$ will increase since it is proportional to $k/\omega$ outside the boundary layer. Finally, the increased $\nu_T$ causes stronger momentum diffusion from the high-momentum region to the low-momentum region, which suppresses the size of the recirculation zone.

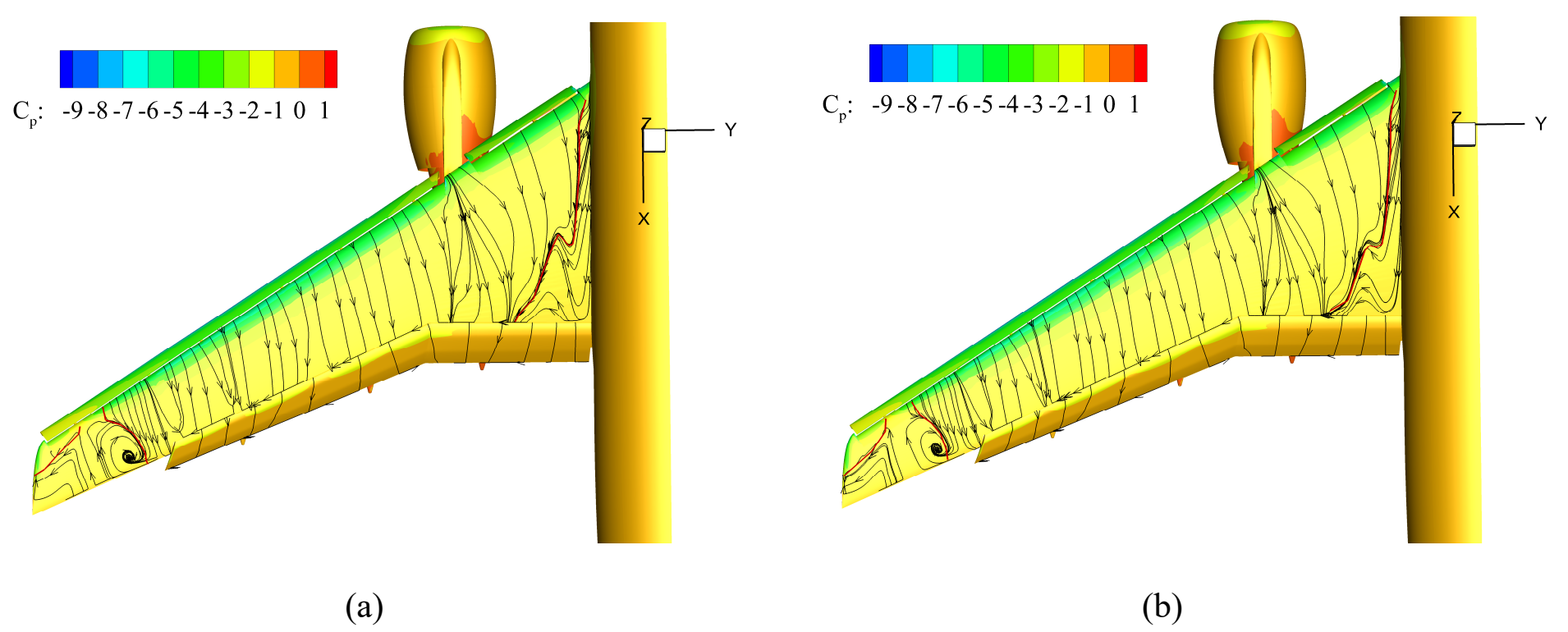

(a) (b)

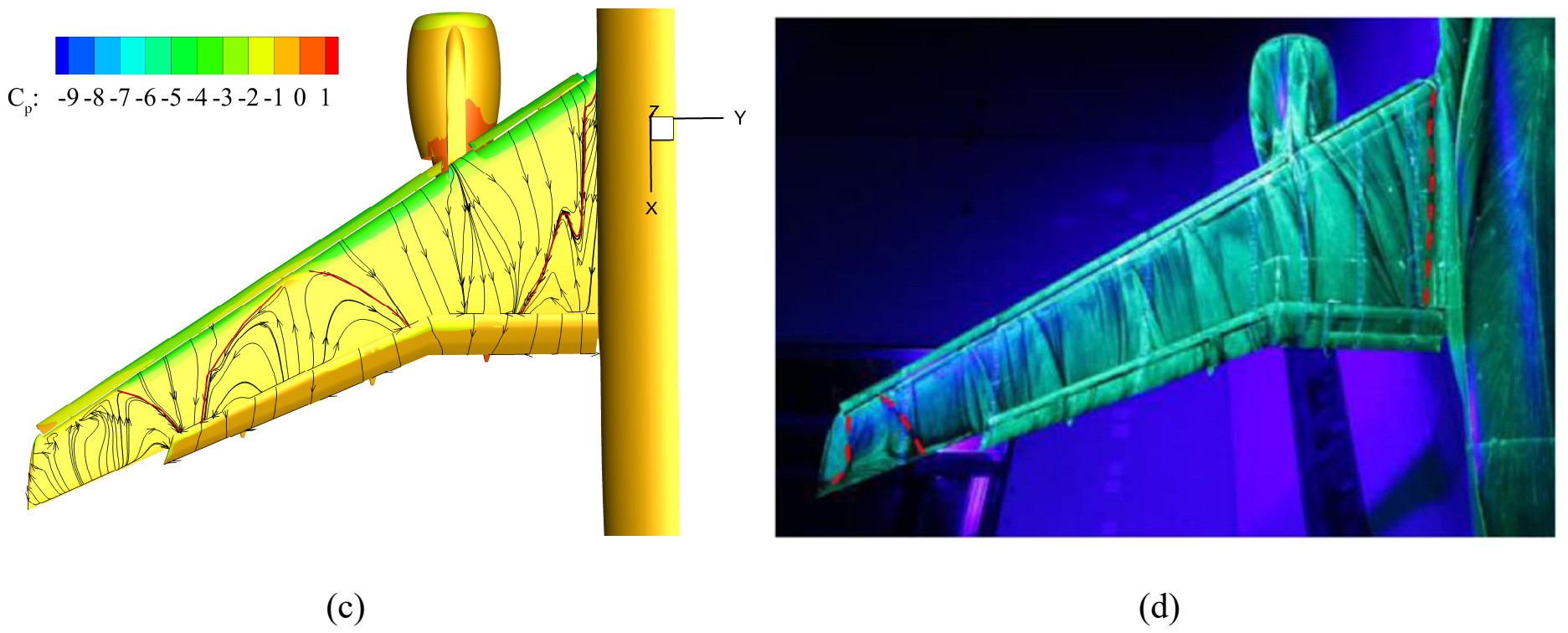

(c) (d)

Figure 34. Surface streamline plot at $AOA = 18.6°$ given by (a) SST-CND model, (b) SST-CND3D model, (c) SST model, (d) Experiment [3][43]

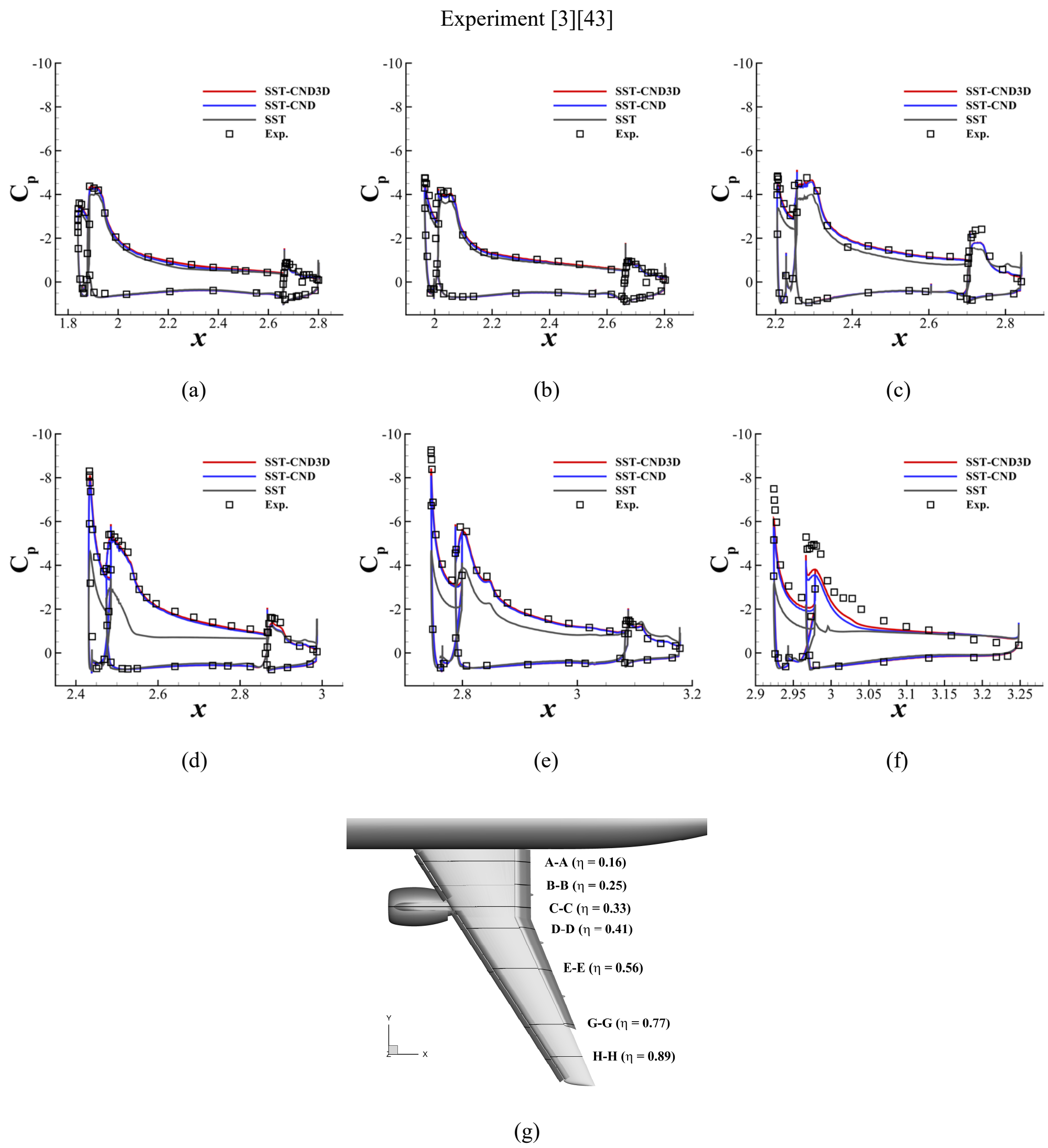

Figure 35. $C_p$ distribution along different sections in the spanwise direction at $AOA = 18.6°$, (a) A-A section, (b) B-B section, (c) D-D section, (d) E-E section, (e) G-G section, (f) H-H section, (g) Location of different sections

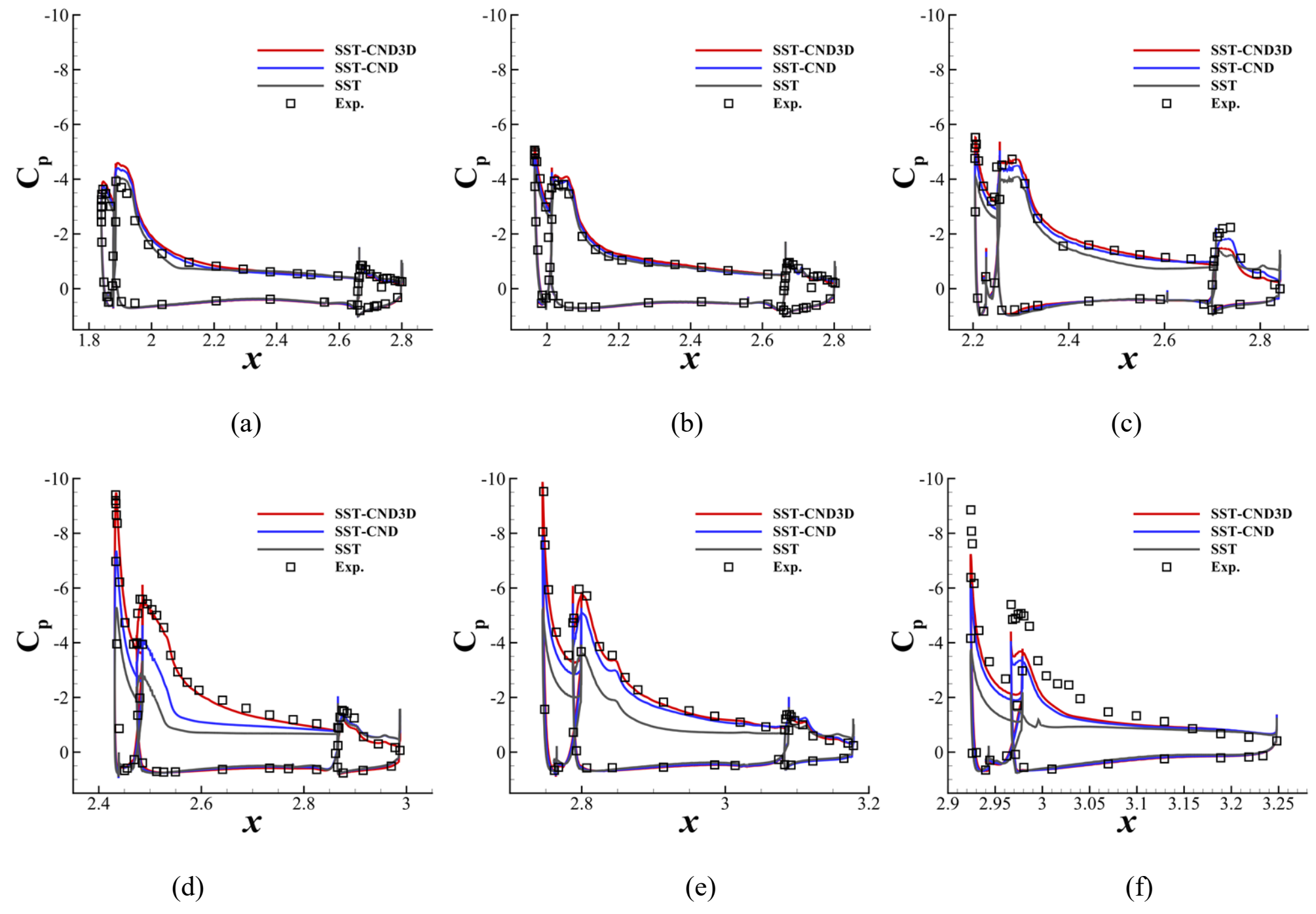

Figure 36. $C_p$ distribution along different sections in the spanwise direction at $AOA = 20.6°$, (a) A-A section, (b) B-B section, (c) D-D section, (d) E-E section, (e) G-G section, (f) H-H section

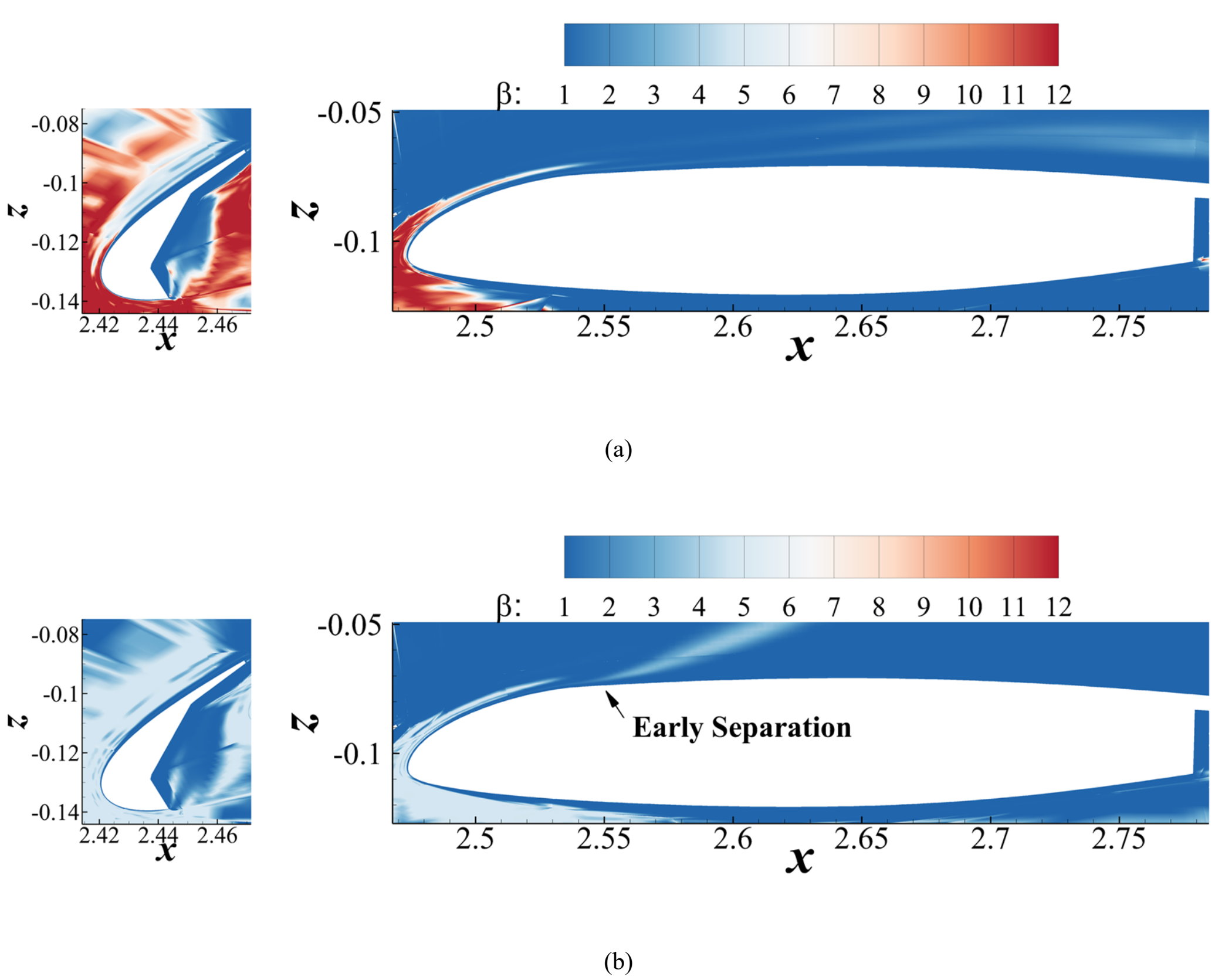

Figure 37. The $\beta$ contours at the leading edge of the wing and slat (E-E section) given by (a) the SST-CND3D model, (b) the SST-CND model

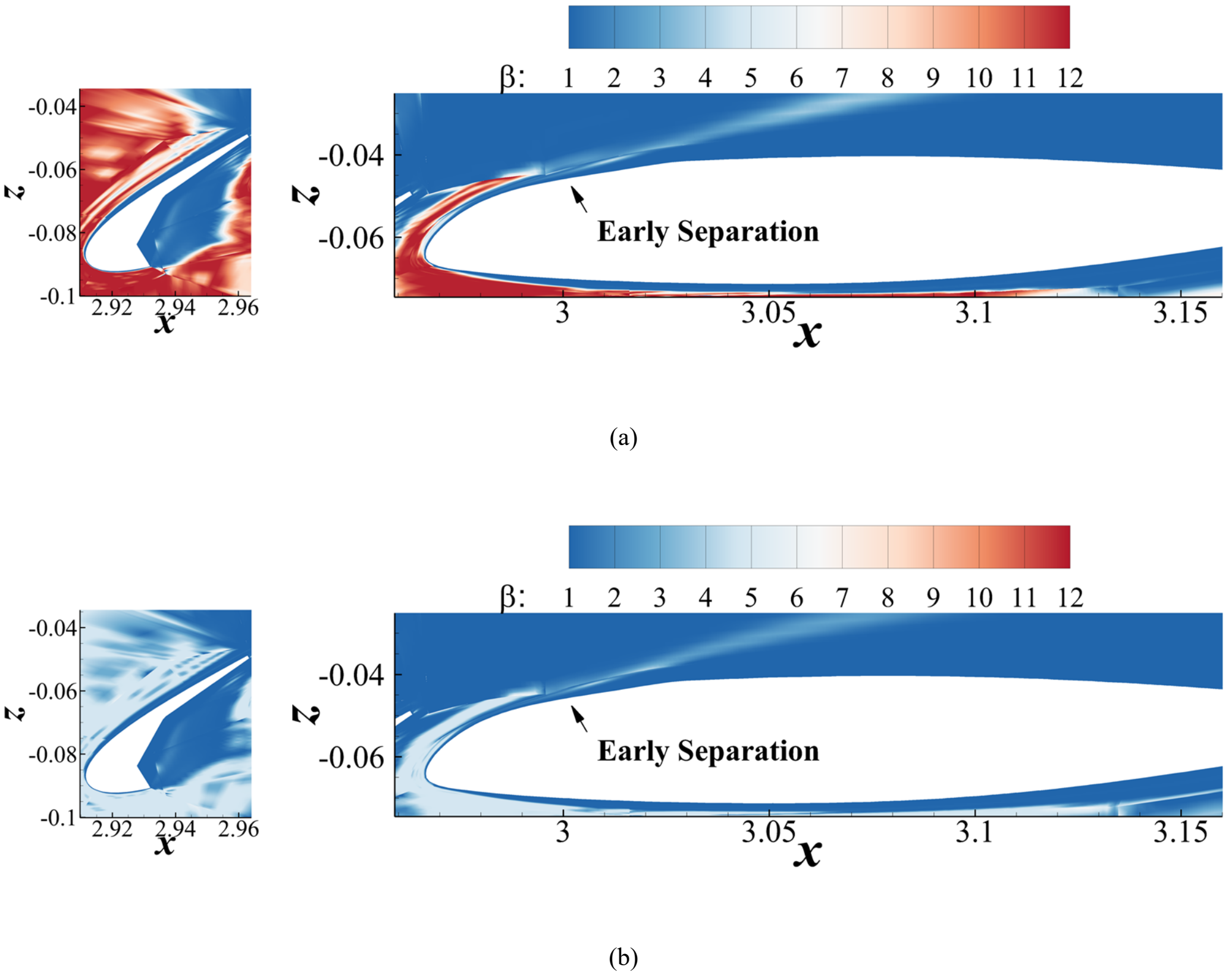

(a)

(b)

Figure 38. The $\beta$ contours at the leading edge of the wing and slat (H-H section) given by (a) the SST-CND3D model, (b) the SST-CND model

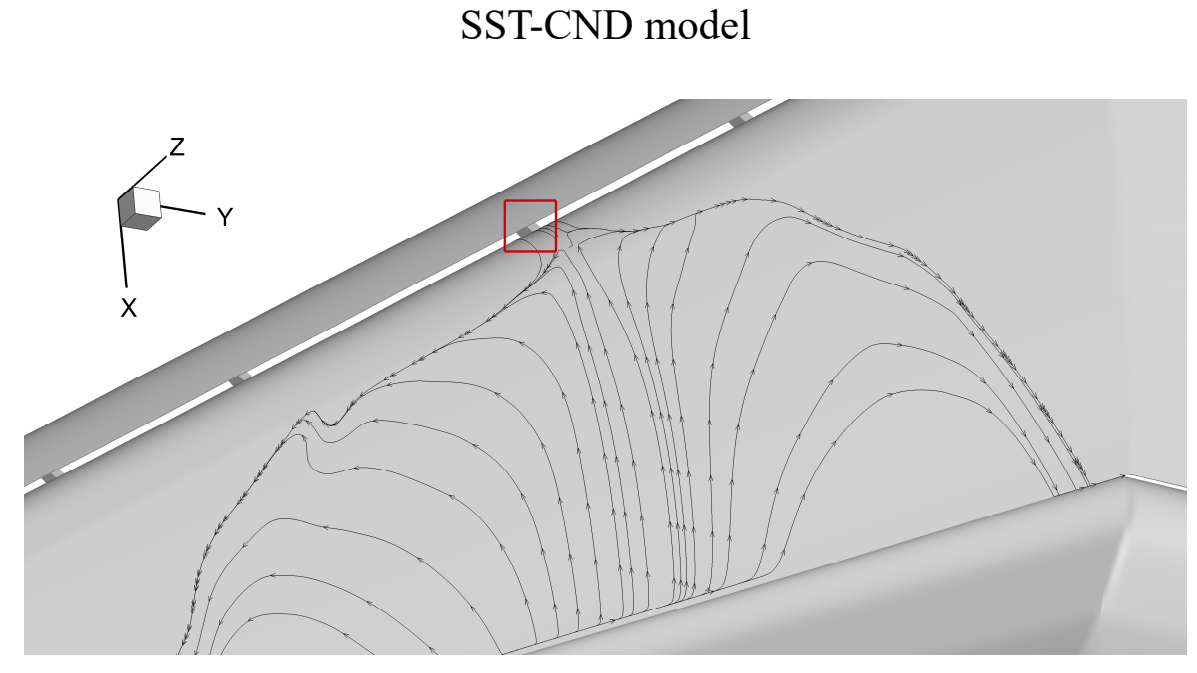

Figure 39. The flow separation induced by the slat track

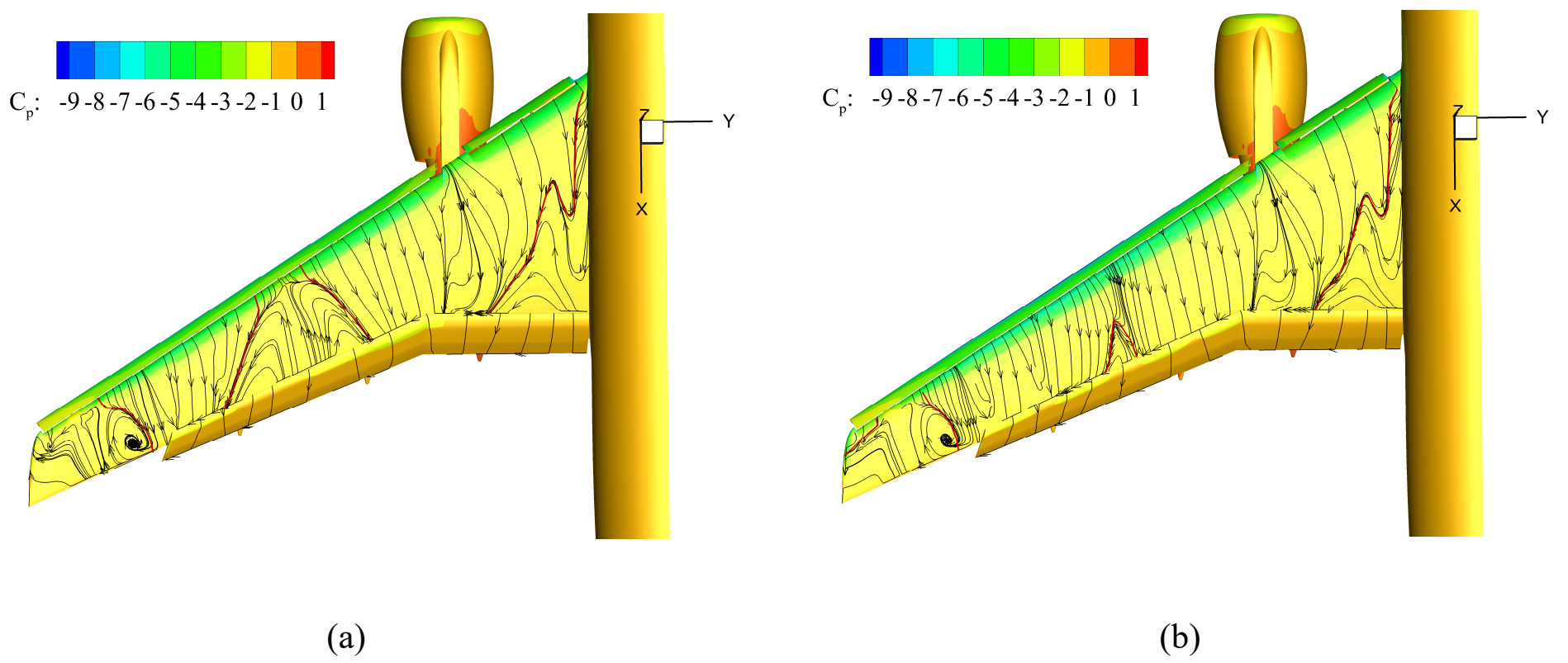

(a) (b)

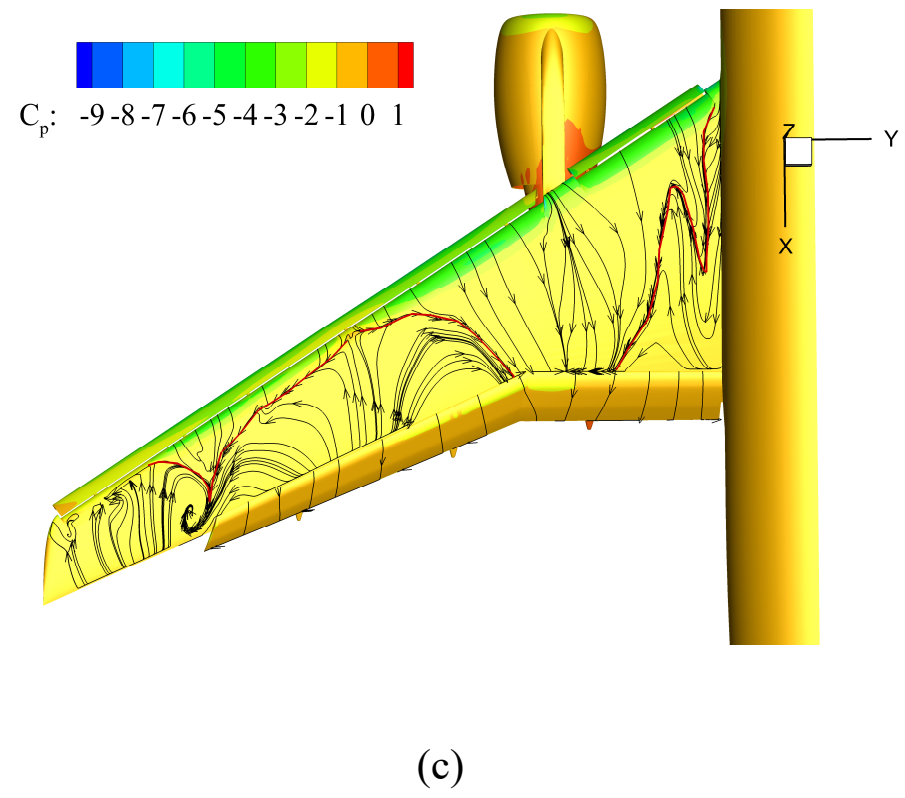

(c)

Figure 40. Surface streamline plot at $AOA = 20.6°$ given by (a) SST-CND model, (b) SST-CND3D model, (c) SST model

The contour of $\beta$ at $y = -1.0\ m$ and the spatial streamlines (projection) are shown in Figure 41. The SST-CND3D model exhibits a larger correction term $\beta$ extending from the wake of the main wing, effectively suppressing the size of the separation bubble.

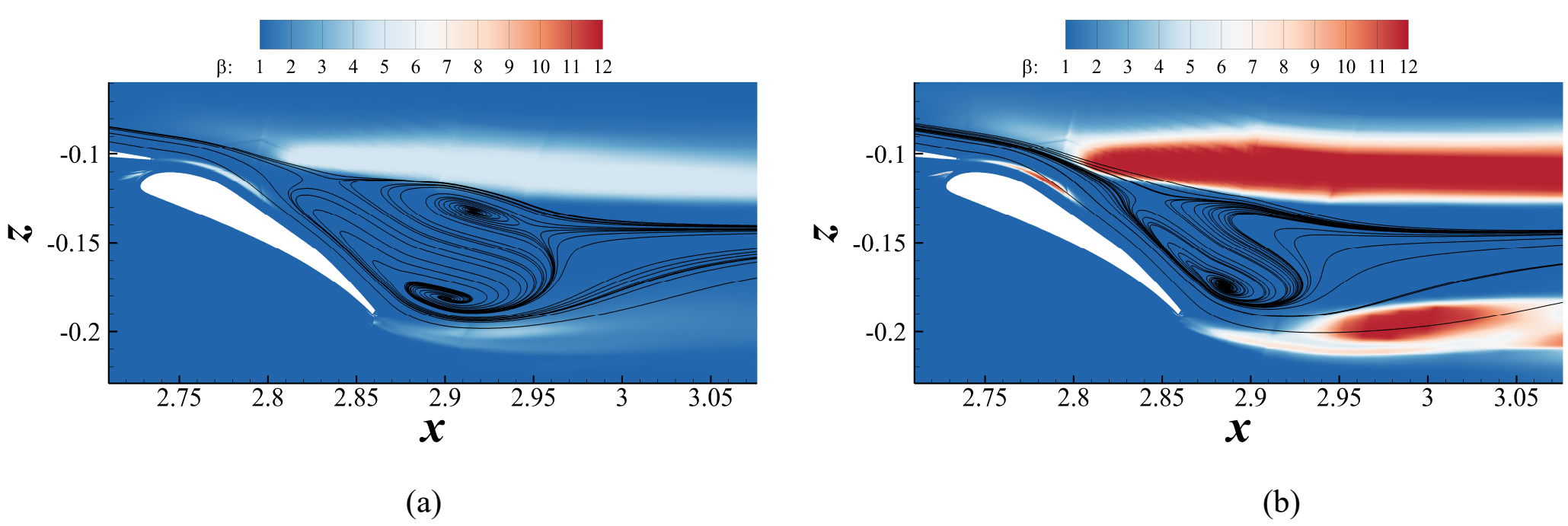

(a) (b)

Figure 41. $\beta$ distribution and the streamline plot (based on $u_x$ and $u_z$) given by different models at $y = -1.0\ m$, $AOA =$ 18.6°, (a) SST-CND model, (b) SST-CND3D model.

Accordingly, the SST-CND3D model outperforms the SST-CND model in the JSM case. The results confirm the effectiveness of the $\beta_{3D}$ correction term constructed through sequential correction and highlight the potential of the SST-CND3D model for application in real-world complex engineering problems.

# 5. Conclusions

This study successfully enhances the SST-CND model's performance in 3D complex flows without retraining the model from scratch or compromising its established accuracy in 2D separated flows employing a sequential correction approach. This approach non-intrusively augments the existing SST-CND model by constructing a new 3D-specific correction term, $\beta_{3D}$, on top of the original formulation. The input features for $\beta_{3D}$ are designed to explicitly vanish in 2D scenarios, ensuring that the correction term primarily accounts for the 3D effects. The explicit algebraic form of $\beta_{3D}$ is determined through a combination of field inversion using 3D cube flow data and symbolic regression. The final expression is selected to ensure that it vanishes in 2D flows. The resultant SST-CND3D model demonstrates identical accuracy to the SST-CND model across all 2D validation cases (hump, flat plate, and multi-element airfoil). The SST-CND3D model in 3D flows achieves superior accuracy compared to the SST-CND model, effectively addressing the underprediction issues of $C_L$ and $AOA_{stall}$ observed in the complex JSM configuration.

This study validates the sequential correction approach as a robust and efficient method for enhancing existing data-driven turbulence models. It enables the incorporation of higher-dimensional data to improve generalization while avoiding the risks and computational costs associated with retraining the entire model. The developed model demonstrates strong potential for application in real-world complex engineering problems.

# Acknowledgement

This work was supported by the National Natural Science Foundation of China (grant numbers 12372288, U23A2069, and 12388101), the National Key Research and Development Program of China (2024YFB4205601), and other national research projects. AI is used to polish the writing of the paper.